\documentclass[11pt]{article}

\usepackage[numbers,square]{natbib}

\usepackage{xcolor}
\usepackage{epsfig}
\usepackage{cancel}
\usepackage{caption}
\usepackage{feynmf} 
\usepackage{tikz-feynman}
\usepackage{amssymb}
\usepackage{amsfonts}
\usepackage{epsf}
\usepackage{rotating}
\usepackage{graphicx}
\usepackage{amsmath}
\usepackage{fancyhdr}
\usepackage{subfigure}
\usepackage{graphics}
\usepackage{pstricks}
\usepackage{color}
\usepackage{frontespizio}
\usepackage{hyperref}
\hypersetup{
	colorlinks,
	citecolor=green,
	filecolor=black,
	linkcolor=blue,
	urlcolor=black
}
\usepackage{type1ec}
\usepackage[T1]{fontenc}
\usepackage{lettrine}
\usepackage{bbold}
\usepackage{calligra}
\usepackage{tikz}
\usepackage{subfigure}
\usepackage{mathrsfs}
\usepackage{curve2e}
\usepackage{setspace}
\usepackage{indentfirst}
\usepackage{emptypage}
\usepackage[babel]{csquotes} 
\usepackage[font=small,labelfont=bf,labelsep=quad]{caption} 
\usepackage{graphicx} 
\usepackage{listings} 
\usetikzlibrary{patterns}
\usepackage{relsize}
\usetikzlibrary{intersections,positioning}
\usetikzlibrary{decorations.pathmorphing,decorations.markings,arrows,positioning}
\usepackage{braket}
\usepackage{mathrsfs}
\usepackage{stackengine}
\usepackage{calc}
\newlength\shlength
\newcommand\xshlongvec[2][0]{\setlength\shlength{#1pt}%
	\stackengine{-5.6pt}{$#2$}{\smash{$\kern\shlength%
			\stackengine{7.55pt}{$\mathchar"017E$}%
			{\rule{\widthof{$#2$}}{.57pt}\kern.4pt}{O}{r}{F}{F}{L}\kern-\shlength$}}%
	{O}{c}{F}{T}{S}}

\newcommand{\sdfrac}[2]{\mbox{\small$\displaystyle\frac{#1}{#2}$}}

\IfFileExists{dsfont.sty}
{\usepackage{dsfont}
	\let\mathbb=\mathds
	\newcommand{\id}{\mathds{1}}}
{\typeout{Package dsfont.sty was not found, using alternative macros.}
	\let\mathds=\mathbb
	\newcommand{\id}{\mbox{1 \kern-.59em {\rm l}}}}

\usepackage{slashed}
\usepackage{units}
\usepackage{setspace}
\renewcommand{\Re}{\textrm{Re}}

\newcommand{\nn}{\nonumber}
\let\a=\alpha   \let\b=\beta   \let\g=\gamma   \let\d=\delta
         
    \let\k=\kappa  \let\l=\lambda  \let\m=\mu
      \let\x=\xi

\let\D=\Delta
\let\d=\delta

\newcommand{\secref}[1]{Section~\ref{#1}}		
\newcommand{\appref}[1]{Appendix~\ref{#1}}		
\renewcommand{\a}{\alpha}

\def\nbox#1#2{\vcenter{\hrule \hbox{\vrule height#2in
			\kern#1in \vrule} \hrule}}
\def\sq{\,\raise.5pt\hbox{$\nbox{.09}{.09}$}\,}
\def\sqb{\,\raise.5pt\hbox{$\overline{\nbox{.09}{.09}}$}\,}

\newcommand{\bea}{\begin{eqnarray}}
\newcommand{\eea}{\end{eqnarray}}
\newcommand{\be}{\begin{equation}}
\newcommand{\ee}{\end{equation}}

\newcommand{\bes}{\begin{subequations}}
	\newcommand{\ees}{\end{subequations}}

\def\nn{\nonumber\\}

\numberwithin{equation}{section}

\usepackage{accents}

\makeatletter
\newcommand{\xLine}[2][]{\ext@arrow 0359\Rightarrowfill@{#1}{#2}}
\makeatother
\xdefinecolor{darkgreen}{RGB}{102, 204, 70}
\xdefinecolor{darkblue}{RGB}{0, 0, 153}

\begin{document}
\begin{center}
\vspace{1.5cm}
{\Large \bf Four-Point Functions in Momentum Space: \\
Conformal Ward Identities in the Scalar/Tensor case\\ }
\vspace{0.3cm}
\vspace{1cm}
{\bf Claudio Corian\`o, Matteo Maria Maglio and Dimosthenis Theofilopoulos\\}
\vspace{1cm}
{\it  Dipartimento di Matematica e Fisica, Universit\`{a} del Salento \\
and  INFN Sezione di Lecce,  Via Arnesano 73100 Lecce, Italy\\}

\end{center}
\begin{abstract}
We derive and analyze the conformal Ward identities (CWI's) of a tensor 4-point function of a generic CFT in momentum space. The correlator involves the stress-energy tensor $T$ and three scalar operators $O$ ($TOOO$). We extend the reconstruction method for tensor correlators from 3- to 4-point functions, starting from the transverse traceless sector of the $TOOO$. We derive the structure of the corresponding CWI's in two different sets of variables, relevant for the analysis of the 1-to-3 (1 graviton $\to$ 3 scalars) and 2-to-2 (graviton + scalar $\to$ two scalars)  scattering processes.  The equations are all expressed in terms of a single form factor. In both cases we discuss the structure of the equations and their possible behaviors in various asymptotic limits of the external invariants. A comparative analysis of the systems of equations for the $TOOO$ and those for the $OOOO$, both in the general (conformal) and dual-conformal/conformal (dcc) cases, is presented. We show that in all the cases the Lauricella functions are homogenous solutions of such systems of equations, also described as parametric 4K integrals of modified Bessel functions. 
 \end{abstract}
\newpage
\section{Introduction} 

The analysis of the conformal constraints in general conformal field theories (CFT's) in momentum space provides new insight into the structure of the corresponding correlators. It allows a direct comparison between  general CFT predictions and those derived within the traditional S-matrix approach - based on the study of scattering amplitudes - widely investigated in a perturbative context. \\
Up to 3-point functions, the conformal Ward identities (CWI) are sufficient to fix all the correlators in terms only of the conformal data, which amount to a set of constants. A similar analysis of higher point functions is far more demanding, since it requires the use of the operator product expansion and the study of conformal partial waves (conformal blocks) associated to a given CFT \cite{Poland:2018epd,Dolan:2000ut}.\\
By turning to momentum space, even the analysis of 3-point functions becomes nontrivial, and one has to proceed with a substantial reformulation of the action of the conformal generators in these new variables, which show the hypergeometric nature of the solutions of the CWI's 
\cite{Coriano:2013jba, Bzowski:2013sza, Bzowski:2019kwd, Bzowski:2018fql} 
\cite{Coriano:2018bsy,Coriano:2018bbe,Maglio:2019grh}. These can all be 
reformulated as systems of partial differential equations (pde's), whose solutions are linear combinations of Appell functions ($F_4$), which are hypergeometric functions of 2 variables. \\
In tensor correlators, by appropriate shifts of the parameters of such solutions, it is possible to solve for all the form factors \cite{Coriano:2018bbe, Coriano:2018bsy} of a given tensorial parameterization. Equivalently, one can map such solutions to parametric integrals (3K integrals) of Bessel functions \cite{Bzowski:2013sza}, which allow to handle the symmetries of a certain correlator quite efficiently. \\
There are four fundamental solutions of a hypergeometric system of pde's generated by the CWI's of a scalar 3-point function, as discussed in \cite{Coriano:2013jba}. Any other solution, obtained by requiring specific symmetries of the correlation function, is built around such a basis \cite{Coriano:2018bsy,Coriano:2018bbe}. This holds also for inhomogenous systems, as illustrated for nontrivial correlators such as the $TJJ$, $TTT$, and so on, where the several form factors appearing in the tensor decomposition can all be determined explicitly in terms of few constants \cite{Bzowski:2013sza}
.\\
As we move to 4-point functions, CWI's cease to provide sufficient information for the complete identification of the corresponding correlators, and it is necessary to define a bootstrap program in momentum space which is consistent with the same CWI's, in analogy with coordinate space. There is hope that, in the near future, also these missing links will soon be solved, allowing for equally valuable, complementary approaches both in momentum and in coordinate space.
\subsection{CFT's and anomalies}
Several parallel studies have widened the goal of this activity, addressing issues such as the use of conformal blocks/CP symmetric blocks (Polyakov blocks) \cite{Isono:2019ihz,Isono:2018rrb,Isono:2019wex} \cite{Chen:2019gka} as well as light-cone blocks \cite{Gillioz:2019lgs, Gillioz:2018mto,Gillioz:2018kwh}, analytic continuations to Lorentzian spacetimes \cite{Bautista:2019qxj} and spinning correlators, just to mention a few, all in momentum space. Related analysis have explored the link to Witten diagrams within the AdS/CFT correspondence \cite{Anand:2019lkt,Albayrak:2019yve}. At the same time, the extension of these investigations to de Sitter space has laid the foundations for new applications in cosmology \cite{Arkani-Hamed:2018kmz,Baumann:2019oyu,Arkani-Hamed:2017fdk,Benincasa:2019vqr,Benincasa:2018ssx} \cite{Kundu:2014gxa} 
\cite{Almeida:2017lrq} and in gravitational waves \cite{Almeida:2019hhx}. Finally, investigations of such correlators in Mellin space \cite{Penedones:2010ue,Fitzpatrick:2011ia} \cite{Gopakumar:2016cpb,Gopakumar:2016wkt} offer a new perspective on the bootstrap program both in flat and in curved space \cite{Sleight:2019mgd,Sleight:2019hfp}, providing further insight into the operatorial structure of a given CFT, and connecting in a new way momentum space and Mellin variables.  \\
Undoubtedly, CWI's play a crucial role in this effort, with widespread applications both at zero and at finite temperature \cite{Ohya:2018qkr}.
 Among all the possible correlators that one may investigate, those containing stress-energy tensors $(T)$  play a special role, due to the presence of the conformal anomaly \cite{Coriano:2012wp}. Analysis of 4-point functions have so far been limited to scalar correlators in flat \cite{Maglio:2019grh} \cite{Bzowski:2019kwd} and curved backgrounds \cite{Arkani-Hamed:2018kmz,Baumann:2019oyu}.
 The level of complexity increases drastically for 3-point functions as soon as one considers correlators containing multiple insertions of stress-energy tensors. Their CWI's, in this case, have to reproduce the correct expression of the conformal anomaly. This introduces significant complications respect to coordinate space where, in general, 
 the issues of the ultraviolet  behaviour at coincident spacetime points of the corresponding operators is not addressed. 
 In coordinate space, the problem has been investigated in few cases - for instance in the $TTT$ case - quite directly, by solving the CWI's separately in their homogenous and inhomogeneous (anomalous) forms, by adding to the homogeneous solution one extra contribution \cite{Osborn:1993cr}. \\
  Such additional contribution amounts to an ultralocal term in the corresponding correlation function, generated when all the coordinates of the operators coalesce \cite{Osborn:1993cr,Erdmenger:1996yc} and reproduced by a variation of the anomaly functional.\\
In this context, studies of such correlators in momentum space find significant guidance from free field theory realizations. For example, direct one-loop computations in classical conformal invariant theories 
(such as massless QED and QCD) indicate that the anomalous breaking of conformal symmetry is associated with the exchange of massless poles \cite{Giannotti:2008cv,Armillis:2009pq,Coriano:2018bsy}. This special feature unifies both conformal and chiral anomalies, as found in supersymmetric studies \cite{Coriano:2014gja}, and it has been shown to be consistent with the solutions of the CWI's  of three point functions, such as the $TJJ$ \cite{Armillis:2009pq,Coriano:2018zdo} and the $TTT$ \cite{Coriano:2018bsy}.
\subsection{Moving to 4-point functions}
The investigation of the CWI's in momentum space that we are going to present is based on the reconstruction method of a tensor correlator starting from its transverse/traceless (tt) sector, formulated for 3-point functions  \cite{Bzowski:2013sza,Bzowski:2018fql,Bzowski:2015yxv}, that here we are going to extend to 4-point functions.  \\
In particular, in \cite{Bzowski:2013sza} a complete approach for the analysis of 3-point functions, up to the $TTT$ case, with three stress-energy tensors, valid for tensor correlators, has been formulated. The reconstruction of the entire correlator from its $tt$ projection involves the identification of a minimal set of form factors in this sector, and it is accompanied by a set of technical steps for re-assembling it in a systematic way. \\
This approach allows to identify primary and secondary CWI's of a tensorial 3-point function, with the former corresponding to second order partial differential equations (pde's) which can be solved independently in terms of a set of arbitrary constants.\\
Primary CWI's are equations involving only form factors of the tt sector and generate, for tensor correlators, inhomogenous systems of pde's of hypergeometric type. Secondary CWI's, on the other end, connect the same form factors to 3-point functions via the corresponding canonical WI's, which impose extra constraints on the constants appearing in the solution of the primary equations. \\
The results that we present extend a previous analysis devoted to the scalar case, involving the $OOOO$ correlator  \cite{Maglio:2019grh}. We will re-investigate the scalar case, by taking a closer look at the structure of the equations and at their asymptotic behaviour. We will remark few additional properties of such correlators and highlight some properties of the asymptotic solutions of such equations, which have not been addressed before. 
This will allow us to gain a more general perspective both on the scalar and the tensor cases, especially in view of possible future extensions of our work to correlators of higher rank.\\
Scalar correlators are characterised only by primary CWI's and are therefore simpler to handle, differently from the $TOOO$ case where both primary and secondary equations are present.\\
In the scalar case the analysis of the conformal constraints will be performed by focusing on a special class of solutions of such equations which are conformal and dual conformal at the same time, derived in \cite{Maglio:2019grh}. These are obtained by imposing a specific condition on the scaling dimensions of the scalar operators, which allow to reduce the CWI's to a hypergeometric system, as in the case of 3-point functions. We have summarised 
their construction in a nutshell in appendix E.

As we move from 3- to 4-point functions, all the equations, primary and secondary, are expressed in terms of 6 invariants, which are the external invariant masses $p_i^2$ and the two Mandelstam invariants $s$ and $t$. As far as we keep the external lines off-shell, and stay away from kinematical points where an invariant is exactly zero, the equations are 
well-defined and it is possible to investigate their structure.  As we are going to show, the selection of a set of specific invariants, compared to others, is particularly beneficial if we intend to uncover the symmetries of the equations and their redundancies under the permutations of the external momenta. \\
A crucial goal of our study is the identification of the asymptotic behaviour of the solutions of such constraints in specific kinematical limits. This may allow, in the near future, to relate results from ordinary 
perturbation theory - in ordinary Lagrangian realizations, at one loop level - to those 
derived from CFT's in the same limits. For instance, in \cite{Coriano:2018bbe,Coriano:2018bsy} it is shown how to match the general solutions of the CWI's for the $TTT$ and $TJJ$ correlators, to free field theories with a specific content of fermions, scalars 
and spin 1 fields. The matching allows to re-express the solutions of such equations in terms of simple one-loop master integrals in full generality, for any CFT.  

\subsection{The search for asymptotic solutions}
For this reason, the search for asymptotic solutions of the CWI's, which acquire a simpler form in such limits, is particularly interesting. It may allow to establish a link with the classical factorization theorems proven in gauge theory amplitudes \cite{Sterman:2014nua}, especially if such CFT methods can be extended to multi-point functions.\\
We will investigate the structure of the equations in two specific limits. The first case that we will address will be the $1\to 3$, where the graviton line of the T is assumed to acquire a large invariant mass $(p_1^2)$ and decays into three scalar lines with small invariants 
$(p_2^2,p_3^2,p_4^2)$, while the remaining invariants $s$ and $t$  are large. We are going to derive some approximate asymptotic solutions of the equations which are separable in the $(p_2^2,p_3^2,p_4^2)$ and $(s,t,u)$ dependence. A similar analysis will be presented in the $2\to 2$ process, where one of the lines of the scalar operators is selected in the initial state together with the graviton line and the remaining scalar lines are in the final state.\\ Our work is organized as follows. \\
After a brief discussion of the conformal and canonical WI's in momentum space, we investigate the structure of the tt sector of the $TOOO$, identifying the symmetry constraints under the permutation of the momenta of the single form factor appearing in this correlator. \\
We then turn to a derivation of the primary and secondary CWI's of this correlator, written in a form which will be useful for the derivation of their asymptotic limits. We describe the orbits of such equations under the symmetry permutations, which allows to identify a subset of independent equations. \\
The analysis is repeated from scratch in the $2\to 2$ case and it is followed by a discussion of the asymptotic limits of such equations, after a brief overview of the approach in the scalar case. \\
We start from the scalar case, discuss the system of scalar equations and discuss its reduction to the dcc case, which can be solved exactly. The asymptotic behaviour of the 
dcc solutions provides an example and a guidance for a more general analysis first of the scalar case, and then of the tensor case, the $TOOO$.
 In our conclusions we present some perspective for further future extensions of our work.

 \section{Ward identities for the $TOOO$ in coordinate and in momentum space} 
 In this section we briefly review the structure of the CWI's in coordinates and momentum space before turning to an analysis of the tensor case.\\
We recall that for scalar correlators of individual scaling dimensions $\Delta_i$
 \begin{equation}
\label{defop}
\Phi(x_1,x_2,\ldots,x_n)=\braket{O_1(x_1)O_2(x_2)\ldots O_n(x_n)}
\end{equation}
with primary scalar operators $O_i$, the special CWI's are given by first order differerential equations 
\begin{equation}
\label{SCWI0}
{\cal K}_{scalar}^\kappa(x_i) \Phi(x_1,x_2,\ldots,x_n) =0,
\end{equation}
with 
\begin{equation}
\label{transf1}
{\cal K}_{scalar}^\kappa(x_i) \equiv \sum_{j=1}^{n} \left(2 \Delta_j x_j^\kappa- x_j^2\frac{\partial}{\partial x_{j,\kappa}}+ 2 x_j^{\kappa} x_j^{\alpha} \frac{\partial}
{\partial x_j^\alpha} \right)
\end{equation}
being the corresponding generator in coordinate space. Denoting with 
\begin{equation}
 \Phi(p_1,\ldots p_{n-1},\bar{p}_n)=\langle O_1(p_1)\ldots O_n(\bar{p}_n)\rangle 
\end{equation}
and 
\begin{equation}
\label{dd1}
{\cal K}_{scalar}^\kappa(p_i)\equiv\sum_{j=1}^{n-1}\left(2(\Delta_j- d)\frac{\partial}{\partial p_{j,\kappa}}+p_j^\kappa \frac{\partial^2}{\partial p_j^\alpha\partial p_j^\alpha} -2 p_j^\alpha\frac{\partial^2}{\partial p_{j,\kappa} \partial p_j^\alpha}\right)
\end{equation}
 the Fourier transform of \eqref{defop} and of \eqref{transf1} respectively, the form of second order differential equations is given by
\begin{equation}
{\cal K}_{scalar}^\kappa(p_i)\Phi(p_1,\ldots p_{n-1},\bar{p}_n)=0,\label{SCWI}
\end{equation}
where we have chosen $\bar{p}_n^\mu=-\sum_{i=1}^{n-1} p_i^\mu$ the n-th momentum, to be the linearly dependent one. These constraints are accompanied by the corresponding dilatation WI's 

\begin{equation}
\phi(\lambda x_i)=\lambda^{-\Delta}\phi(x_i), 
\end{equation}
which reduce to the form 
\begin{equation}
\label{scale12}
D(x_i)
\Phi(x_1,\ldots x_n)=0,
\end{equation}
with the (Euler) operator $D(x_i)$ given by
\begin{equation}
\label{scale11}
D(x_i)\equiv\sum_{i=1}^n\left( x_i^\alpha \frac{\partial}{\partial x_i^\alpha} +\Delta_i\right).
\end{equation}

In momentum space, the dilatation WI is then given by 
\begin{equation}
D(p_i) \Phi(p_1\ldots \bar{p}_n)=0,
\end{equation}
where
\begin{equation}
D(p_i)\equiv\sum_{i=1}^{n-1}  p_i^\alpha \frac{\partial}{\partial p_i^\alpha} + \Delta',
\end{equation}
with the overall scaling in momentum space being given by \cite{Coriano:2018bbe}
\begin{equation}
\Delta'\equiv \left(-\sum_{i=1}^n \Delta_i +(n-1) d\right)=-\Delta_t +(n-1) d.
\end{equation}
In the expression above, $\Delta_t=\sum_{i=1}^4 \Delta_i$ denotes the total scaling in coordinate space, while the same scaling in momentum space is associated with $\Delta'$ as 

\begin{equation}
\phi(\lambda p_1\ldots \lambda \bar{p}_n)=\lambda^{-\Delta'}\phi(p_1\ldots \bar{p}_n).
\end{equation}
Coming to the tensor case, we recall that the infinitesimal conformal transformation $x^\mu\to x^\mu + v^\mu(x)$ for the stress-energy tensor 
with

\be
v_{\mu}(x)=b_\mu x^2 -2 x_\mu b\cdot x,
\ee
defined in terms of a generic parameter $b_\mu$, and a scaling factor
\be
\Omega= 1- \sigma +\ldots 
\ee
with  $\sigma=-2 b\cdot x$, can be expressed as a local rotation times a rescaling $\Omega$
\be
\label{rot1}
R^\mu_\alpha=\Omega \frac{\partial x'^\mu}{\partial x^\alpha}, 
\ee
and the action on the stress-energy tensor is simply given by
\bea
T'^{\mu\nu}(x')&=&\Omega^{\Delta_T} R^\mu_\alpha R^\nu_\beta T^{\alpha\beta}(x).
\label{trans}
 \eea
$R$ can be expanded around the identity as 
\be
R=\mathbf{ 1  } + \left[\mathbf{\epsilon}\right] +\ldots
\ee
with an antisymmetric matrix $\left[\epsilon\right]$, which we can re-express in terms of antisymmetric parameters 
($\tau_{\rho\sigma}$) and $1/2 \,d\, (d-1)$ generators $\Sigma_{\rho\sigma}$ of $SO(d)$ as 
\bea
\left[\epsilon\right]_{\mu\alpha}&=&\frac{1}{2} \tau_{\rho\sigma}\left(\Sigma_{\rho\sigma}\right)_{\mu\alpha}\nonumber \\
\left(\Sigma_{\rho\sigma}\right)_{\mu\alpha}&=&\delta_{\rho\mu}\delta_{\sigma\alpha}-\delta_{\rho\alpha}\delta_{\sigma\mu}
\eea
\be
R_{\mu\alpha}= \delta_{\mu\alpha} + \tau_{\mu\alpha}=\delta_{\mu\alpha} + \frac{1}{2}\partial_{[\alpha }v_{\mu]},
\ee
where $ \partial_{[\alpha }v_{\mu]}\equiv
\partial_{\alpha }v_{\mu}-\partial_{\mu }v_{\alpha}$.\\
One derives from \eqref{trans} the infinitesimal transformation 
\be
\delta T^{\mu\nu}(x)=-(b^\alpha x^2 -2 x^\alpha b\cdot x )\, \partial_\alpha  T^{\mu\nu}(x)   - \Delta_T \sigma T^{\mu\nu}(x)+
2(b_\mu x_\alpha- b_\alpha x_\mu)T^{\alpha\nu} + 2 (b_\nu x_\alpha -b_\alpha x_\nu)\, T^{\mu\alpha}(x).
\ee
It is then quite straightforward to obtain the expression of the special CWI for the correlator 
\be
\Gamma^{\mu\nu}(x_1,x_2,x_3,x_4)\equiv \langle T^{\mu\nu}(x_1)O(x_2)O(x_3)O(x_4) \rangle
\ee
in the form 
\bea
{\cal K}^\kappa \Gamma^{\mu\nu}(x_1,x_2,x_3,x_4) 
&=&  {\cal K}^{ \kappa}_{scalar}(x_i) \Gamma^{\mu\nu}(x_1,x_2,x_3,x_4) 
 + 2 \left(  \delta^{\mu\kappa} x_{1\rho} - \delta_{\rho}^{\kappa }x_1^\mu  \right)\Gamma^{\rho \nu}(x_1,x_2,x_3,x_4)\nn\\ 
&& + 2 \left(  \delta^{\nu\kappa} x_{1\rho} - \delta_{\rho}^{\kappa }x_1^\nu  \right)\Gamma^{\mu\rho}(x_1,x_2,x_3,x_4)\nonumber
=0,
 \eea
 where the first contribution denotes the scalar part and the last two contributions the spin part, which are trivially absent in the case of a scalar correlator. \\
 The transition to momentum space of such equations has been discussed in \cite{Coriano:2018bbe}, to which we will refer for further details, and the action of $\mathcal{K}^\kappa$ can be summarized by the expression 

 \begin{align}
&\sum_{j=1}^{3}\left[2(\Delta_j-d)\sdfrac{\partial}{\partial p_j^\k}-2p_j^\a\sdfrac{\partial}{\partial p_j^\a}\sdfrac{\partial}{\partial p_j^\k}+(p_j)_\k\sdfrac{\partial}{\partial p_j^\a}\sdfrac{\partial}{\partial p_{j\a}}\right]\braket{{T^{\mu_1\nu_1}(p_1)\,O(p_2)\,O( p_3)}O(\bar{p_4}}\notag \\
&\qquad + {\cal K}^\kappa_{spin}\braket{{T^{\mu_1\nu_1}(p_1)\,O(p_2)\,O( p_3)}O(\bar{p_4}}=0,\notag\\
\label{SCWTJJ}
\end{align}
where we have defined the spin part of ${\cal K}$ in momentum space as

\bea
{\cal K}^\kappa_{spin}\braket{{T^{\mu_1\nu_1}(p_1)\,O(p_2)\,O( p_3)}O(\bar{p_4}}
&\equiv& 4\left(\d^{\k(\mu_1}\sdfrac{\partial}{\partial p_1^{\a_1}}-\delta^{\k}_{\alpha_1}\delta^{\l(\mu_1}\sdfrac{\partial}{\partial p_1^\l}\right)\braket{{T^{\nu_1)\alpha_1}(p_1)\,O(p_2)\,O p_3)}O(\bar{p}_4)}\nn
\eea
(symmetrization is normalized with an overall factor $1/2$). \\
In the previous expression we have taken $p_4$ as a dependent momentum ($p_4\to \overline{p}_4$), which requires an implicit differentiation if we take $p_1,p_2$ and $p_3$ as independent momenta. The equations can be projected onto the three independent 
momenta, giving scalar equations which can be re-expressed in terms of all the scalar invariants parameterizing the form factors. The hypergeometric character of the 3-point functions, as well as for 4-point functions (for the dcc solutions), emerges after such reduction of the equations to a scalar form \cite{Coriano:2013jba}\cite{Bzowski:2013sza} \cite{Coriano:2018bsy,Coriano:2018bbe}. \\ 
For this purpose, we recall that $F_4$, Appell's 4th hypergeometric function, which is the only special function appearing in the solution, is defined by the series 
\begin{equation} 
\label{f4}
F_4(\alpha,\beta,\gamma, \gamma';x,y)=\sum_{m, n=0}^{\infty}
\frac{(\alpha)_{m+n}(\beta)_{m+n}}{(\gamma)_{m}(\gamma')_n m! n!}x^m y^n
\end{equation}
with the (Pochhammer) symbol $(\alpha)_{k}$ given by
\begin{equation}
(\alpha)_{k}=\frac{\Gamma(\alpha+k)}{\Gamma(\alpha)}=\alpha(\alpha+1)\dots(\alpha+k-1).\label{Pochh}
\end{equation}
Such function appears in the solution of the CWI's of the scalar 3-point correlator 
\cite{Coriano:2013jba}
\begin{equation} 
 \Phi(q_1,q_2,q_3)=\langle O(q_1)O(q_2) O(q_3)\rangle,
\end{equation}
given by
\begin{equation} 
K_1 \Phi(q_1,q_2,q_3)=0\qquad K_2 \Phi(q_1,q_2,q_3)=0\qquad  K_3 \Phi(q_1,q_2,q_3)=0, 
\label{eqsh}
\end{equation}
where the $K_i$ are given by 
\begin{equation}
K_j=\frac{\partial^2}{\partial p_j^2}+\frac{(d-2\Delta_j+1)}{p_j}\frac{\partial}{\partial p_j},\qquad j=2,3,4\,,
\label{kappa}
\end{equation}
and \eqref{eqsh} can be combined into two equations (see appendix D)
\begin{equation}
K_{13}\Phi=0\qquad K_{23}\Phi=0, \qquad \textrm{where} \qquad K_{ij}=K_i -K_j.
\label{ipergio}
\end{equation}

 In this case, following the discussion of \cite{Coriano:2013jba,Coriano:2018bbe}, they can be solved by the linear combination of Appell functions
\begin{equation}
\Phi(q_1,q_2,q_3)=\big(q_3^2\big)^{\frac{\Delta_t}{2} -\frac{3}{2}d -4} \sum_{a,b} c(a,b,\Delta)\,x^a y^b \,F_4(\alpha(a,b), \beta(a,b); \gamma(a), \gamma'(b); x, y), 
\label{compact}
\end{equation}
where here 
\begin{equation}
 x=q_1^2/q_3^2, \qquad y=q_2^2/q_3^2 
\label{qquad}
 \end{equation}
 are quadratic ratios of momenta, expressed in terms of a pivot, which in this case is $q_3$. 
 The pivot is arbitrary among the three momenta, and changes in the pivot are associated to analytic continuations of the variables \cite{Coriano:2013jba}. 
 In \eqref{compact} we are assuming the same scaling dimension for the three scalars operators 
 ($\Delta_i=\Delta,\, i=1,2,3$). The expressions of $\alpha(a,b),\beta(a,b), \gamma(a), \gamma'(b)$ take the form
\begin{align}
&\alpha(a,b)= a + b + \frac{d}{2} -\frac{\Delta}{2},&& \beta(a,b)= a + b + \frac{d}{2} -\frac{3\Delta}{2}\notag\\
&\gamma(a)=2 a +\frac{d}{2} -\Delta + 1,&&\gamma'(b)=2 b +\frac{d}{2} -\Delta + 1 \label{cons2}
\end{align}
where the $(a,b)$ run on 4 pairs of indices $(a_i,b_j)$ $(i,j,=0,1)$
\begin{align}
\label{FuchsianPoint}
&a_0=0, \qquad a_1=\Delta- \frac{d}{2},\notag\\
&b_0=0, \qquad b_1=\Delta-\frac{d}{2}.
\end{align}
They are identified by the condition that an ans\"azt based on the ratios of momenta $x$ and $y$ is free of non-analytic terms at x=0, y=0 (i.e. $\sim 1/x, 1/y$), which need to vanish \cite{Coriano:2018bbe}. 
Notice that the coefficients $c(a,b,\Delta)$ are not all independent, but they need to satisfy some symmetry constraints. Only a single overall constant appears in the general solution \cite{Coriano:2013jba,Coriano:2018bbe}. \\
Equivalently, they can be written down as an integral of three Bessel functions (3K integral), \begin{equation}
I_{\a\{\b_1,\b_2,\b_3\}}(q_1,q_2,q_3)=C\int_0^\infty\,dx\,x^\a\,(q_1)^{\b_1}\,(q_2)^{\b_2}\,(q_3)^{\b_3}\,K_{\b_1}(q_1\,x)\,K_{\b_2}(q_2\,x)\,K_{\b_3}(q_3\,x), 
\end{equation}
in the form \cite{Bzowski:2013sza}
\begin{equation} 
\Phi(q_1,q_2,q_3)=C I_{{d/2-1}\{\Delta_1-d/2,\Delta_2-d/2,\Delta_3-d/2\}}(q_1,q_2,q_3)
\end{equation}
with $\alpha=d/2-1$ and $\beta_i=\Delta_i-d/2$.

In the next sections, we are going to derive the explicit form of the CWI's for the $TOOO$, extending the approach of \cite{Bzowski:2013sza} from 3- to 4-point functions. Together with the conformal constraints, we need to impose on the correlator also the canonical WI's, which we are now going to derive.

\subsection{Conservation and Trace Ward Identities}
For this goal, we start from the generating functional

\begin{equation}
Z[\phi_0,g^{\mu \nu}]=\int  \mathcal{D} \Phi \mathrm{exp}\big(-S_{CFT}[\phi,g^{\mu \nu}]-\sum_i \int d^d x \sqrt{g}\phi_0^{j}O_j \big),
\end{equation}
dependent on the background metric $g_{\mu \nu}$ and the classical source $\phi_0(x)$ coupled to the scalar operator $O(x)$, with the 1-point functions given by 

\begin{align}
&\langle T^{\mu \nu}(x) \rangle=\frac{2}{\sqrt{g(x)}} \frac{\delta Z}{\delta g_{\mu \nu}},\\  
&\langle O_j (x) \rangle =-\frac{1}{\sqrt{g(x)}}\frac{\delta Z}{\delta \phi_0^j(x)}.
\end{align}
In our case, in order to avoid some bulky notation, we consider only one type of scalar operator, with a unique scaling dimension $\Delta$. We will present the derivations of all the conformal and canonical WI's in this specific case. In section \ref{gen} we will then provide  the expression of the same equations for general distinct $\Delta_i$'s, which can be obtained by a very similar procedure, as in the equal scaling case. \\
To get the transverse and trace Ward Identities, we require that the generating functional Z is invariant under diffeomorphisms and Weyl transformations respectively, which gives
\begin{align}
\nabla_{\nu} \langle T^{\mu \nu} (x_1) \rangle +\partial^{\mu}\phi_0 \cdot \langle O(x_1) \rangle =0,\\
g_{\mu \nu}\langle T^{\mu \nu} (x_1) \rangle+(d-\Delta)\phi_0 \langle O(x_1) \rangle =0.
\end{align}
The WI's for the  $\langle T^{\mu_1 \nu_1}(\bold{p_1})O(\bold{p_2})O(\bold{p_3})O(\bold{p_4})\rangle $ can be derived by taking three variations of the above identities with respect to the source $\phi_0$ of the scalar operator. At the end, by imposing the flat limit $g_{\mu \nu}=\delta_{\mu \nu}, \nabla_{\nu}=\partial_{\nu}$, turning off the sources ($\phi_0=0$) and using the definitions
\begin{align}
&\langle T^{\mu \nu} (x_1) O(x_2)O(x_3)O(x_4)\rangle=\frac{-2}{\sqrt{g(x_1)}\dots\sqrt{g(x_4)}}\frac{\delta^4 Z}{\delta g_{\mu \nu}(x_1)\delta \phi_0(x_2)\delta \phi_0(x_3)\delta \phi_0(x_4)},\\
&\langle O(x_1)O(x_2)O(x_3) \rangle=\frac{-1}{\sqrt{g(x_1)}\sqrt{g(x_2)}\sqrt{g(x_3)}}\frac{\delta^3Z}{\delta \phi_0(x_3)\delta \phi_0(x_2)\delta \phi_0(x_1)},
\end{align}
 the conservation WI gives the constraint
\begin{equation}\label{TransCord}
\begin{split}
\partial_{\nu}\langle T^{\mu \nu} (x_1)O(x_2)O(x_3)O(x_4) \rangle&=\partial^{\mu}\delta^{(d)}(x_1-x_2)\langle O(x_1)O(x_3)O(x_4) \rangle
\\
&\hspace{-2cm}+\partial^{\mu}\delta^{(d)}(x_1-x_3)\langle O(x_1)O(x_2)O(x_4) \rangle+\partial^{\mu}\delta^{(d)}(x_1-x_4)\langle O(x_1)O(x_2)O(x_3) \rangle,
\end{split}
\end{equation}
while the trace WI gives
\begin{align}\label{TraceCoord}
\delta_{\mu \nu} \langle T^{\mu \nu} (x_1)O(x_2)O(x_3))(x_4) \rangle&=(d-\Delta)\delta^{(d)}(x_1-x_2)\langle O(x_1)O(x_3)O(x_4) \rangle \notag\\
&\hspace{-3cm}+(d-\Delta)\delta^{(d)}(x_1-x_3)\langle O(x_1)O(x_2)O(x_4) \rangle +(d-\Delta)\delta^{(d)}(x_1-x_4)\langle O(x_1)O(x_2)O(x_3) \rangle.
\end{align}
The expressions of \eqref{TransCord} and \eqref{TraceCoord} in momentum space can be obtained  by a Fourier transform and are explicitly given by
\begin{equation}\label{TransMom}
\begin{split}
\delta^{(d)}\left(\sum_{i=1}^{4}\textbf{p}_i \right) p_{1\nu}\langle T^{\mu \nu} (\textbf{p}_1)O(\textbf{p}_2)O(\textbf{p}_3)O(\textbf{p}_4) \rangle=&-\Bigg( p_2^{\mu}\langle O(\textbf{p}_1+\textbf{p}_2)O(\textbf{p}_3)O(\textbf{p}_4) \rangle\\
&\hspace{-4cm}+p_3^{\mu}\langle O(\textbf{p}_1+\textbf{p}_3)O(\textbf{p}_2)O(\textbf{p}_4) \rangle+p_4^{\mu}\langle O(\textbf{p}_1+\textbf{p}_4)O(\textbf{p}_2)O(\textbf{p}_3) \rangle\Bigg)\delta^{(d)} \left(\sum_{i=1}^{4}\textbf{p}_i \right),
\end{split}
\end{equation}
and
\begin{equation}\label{TraceMom}
\begin{split}
\delta^{(d)}\left(\sum_{i=1}^{4}\textbf{p}_i \right)\delta_{\mu \nu}\langle T^{\mu \nu} (\textbf{p}_1)O(\textbf{p}_2)O(\textbf{p}_3)O(\textbf{p}_4) \rangle=&(d-\Delta)\Bigg( \langle O(\textbf{p}_1+\textbf{p}_2)O(\textbf{p}_3)O(\textbf{p}_4) \rangle\\
&\hspace{-3cm}+\langle O(\textbf{p}_1+\textbf{p}_3)O(\textbf{p}_2)O(\textbf{p}_4) \rangle +\langle O(\textbf{p}_1+\textbf{p}_4)O(\textbf{p}_2)O(\textbf{p}_3) \rangle\Bigg)\delta^{(d)} \left(\sum_{i=1}^{4}\textbf{p}_i \right),
\end{split}
\end{equation}
where on the right hand side of the equations appear only scalar 3-point functions.
 We will insert a bar over a momentum variable to indicate that it is treated as a dependent one. In the following we are going to make two separate choices of dependent momenta, respectively $\bar{p}_1$ and $\bar{p}_4$. If we choose $\bar{p}_1$ as the dependent momentum, the WI's take the form 
\begin{subequations}
	\begin{align}
	\bar{p}_{\mu_1}\,\braket{T^{\mu_1\nu_1}(\mathbf{\bar{p}}_1)\,O(\mathbf{p}_2)\,O(\mathbf{p}_3)\,O(\mathbf{p}_4)}&=-p_2^{\nu_1}\braket{O(\mathbf{p}_3+\mathbf{p}_4)\,O(\mathbf{p}_3)\,O(\mathbf{p}_4)}\notag\\[1.3ex]
	&\hspace{-3cm}-p_3^{\nu_1}\braket{O(\mathbf{p}_2+\mathbf{p}_4)\,O(\mathbf{p}_2)\,O(\mathbf{p}_4)}-p_4^{\nu_1}\braket{O(\mathbf{p}_2+\mathbf{p}_3)\,O(\mathbf{p}_2)\,O(\mathbf{p}_3)},\\[2ex]
	\delta_{\mu_1\nu_1}\,\braket{T^{\mu_1\nu_1}(\mathbf{\bar{p}}_1)\,O(\mathbf{p}_2)\,O(\mathbf{p}_3)\,O(\mathbf{p}_4)}&=(d-\Delta)\Big[\braket{O(\mathbf{p}_3+\mathbf{p}_4)\,O(\mathbf{p}_3)\,O(\mathbf{p}_4)}\notag\\[1.3ex]
	&\hspace{-3cm}+\braket{O(\mathbf{p}_2+\mathbf{p}_4)\,O(\mathbf{p}_2)\,O(\mathbf{p}_4)}+\braket{O(\mathbf{p}_2+\mathbf{p}_3)\,O(\mathbf{p}_2)\,O(\mathbf{p}_3)}\Big],
	\end{align}\label{ConsWI}
\end{subequations}
and for $\bar{p}_4$
\begin{subequations}
	\begin{align}
	p_{1,\mu_1}\langle T^{\mu_1 \nu_1}(\bold{p_1})O(\bold{p_2})O(\bold{p_3})O(\mathbf{\bar{p}_4})\rangle=&+p_1^{\nu_1}\braket{ O(\bold{p_2})O(\bold{p_3})O(\mathbf{p_1+\bar{p}_4})}\notag \\&-p_2^{\nu_1}\Big(\langle O(\bold{p_1+p_2})O(\bold{p_3})O(\mathbf{\bar{p}_4})\rangle-\langle O(\bold{p_2})O(\bold{p_3})O(\mathbf{p_1+\bar{p}_4}) \rangle\Big)\notag \\&-p_3^{\nu_1}\Big(\langle O(\bold{p_2})O(\bold{p_1+p_3})O(\mathbf{\bar{p}_4})\rangle-\langle O(\bold{p_2})O(\bold{p_3})O(\mathbf{p_1+\bar{p}_4}) \rangle\Big),\\
	\delta_{\mu_1\nu_1}\langle T^{\mu_1 \nu_1}(\bold{p_1})O(\bold{p_2})O(\bold{p_3})O(\mathbf{\bar{p}_4})\rangle=&(d-\Delta)\Big[\langle O(\textbf{p}_1+\textbf{p}_2)O(\textbf{p}_3)O(\mathbf{\bar{p}_4}) \rangle+\langle O(\textbf{p}_1+\textbf{p}_3)O(\textbf{p}_2)O(\mathbf{\bar{p}_4}) \rangle \notag \\&+\langle O(\textbf{p}_1+\mathbf{\bar{p}_4})O(\textbf{p}_2)O(\textbf{p}_3) \rangle\Big].
	\end{align}\label{ConsWI22}
\end{subequations} 
The left hand sides of these equations will be related to the form factor identified from the $tt$ sector.
 
\section{The reconstruction method from 3- to 4-point functions and the $TOOO$}
Following \cite{Bzowski:2013sza}, we consider the four point correlation function formed by a stress-energy tensor $T^{\mu\nu}$ and three scalar operators $O(p_i)$ of the same kind and with the same scaling dimensions. We define
\begin{align}
p_i=\sqrt{\mathbf{p}_i^2},\quad s=\sqrt{(\mathbf{\bar{p}}_1+\mathbf{p}_2)^2}=\sqrt{(\mathbf{p}_3+\mathbf{p}_4)^2},\quad t=\sqrt{(\mathbf{p}_2+\mathbf{p}_3)^2},\quad u=\sqrt{(\mathbf{p}_2+\mathbf{p}_4)^2},
\end{align}
and introduce the $tt$ $(\Pi)$ and local $(\Sigma)$ projectors 
\begin{align}
\Pi^{\mu\nu}_{\alpha\beta}(\mathbf{p})&=\frac{1}{2}\left(\pi^{\mu}_{\alpha}(\mathbf{p})\pi^{\nu}_{\beta}(\mathbf{p})+\pi^{\mu}_{\beta}(\mathbf{p})\pi^{\nu}_{\alpha}(\mathbf{p})\right)-\frac{1}{d-1}\pi^{\mu\nu}(\mathbf{p})\pi_{\alpha\beta}(\mathbf{p})\\
\Sigma^{\mu\nu}_{\alpha\beta}(\mathbf{p})&=\delta^{(\mu}_{\alpha}\delta^{\nu)}_{\beta}-\Pi^{\mu\nu}_{\alpha\beta}(\mathbf{p})=\frac{1}{p^2}\left[2\,p_{\scriptsize{\raisebox{-0.9ex}{$(\beta$}}}\,\delta^{(\mu}_{\scriptsize{\raisebox{-0.7ex}{$\alpha)$}}} p^{\scriptsize{\raisebox{0.5ex}{$\nu)$}}}-\frac{p_\alpha p_\beta}{(d-1)}\left(\delta^{\mu\nu}+(d-2)\frac{p^\mu p^\nu}{p^2}\right)\right]+\frac{1}{d-1}\pi^{\mu\nu}(\mathbf{p})\delta_{\alpha\beta}\label{Sigma}.
\end{align}
The stress-energy tensor is decomposed in its transverse traceless $(tt)$ and local parts in the form 

\begin{equation}
\label{loca1}
T^{\mu\nu}=t^{\mu\nu} + t_{loc}^{\mu\nu}
\end{equation}
with 
\begin{align}
\label{loca2}
t_{loc}^{\mu\nu}(p)&=\frac{p^{\mu}}{p^2}Q^\nu + \frac{p^{\nu}}{p^2}Q^\mu -
\frac{p^\mu p^\nu}{p^4} Q +\frac{\pi^{\m\nu}}{d-1}(T - \frac{Q}{p^2})\nn
&=\Sigma^{\mu\nu}_{\alpha\beta} T^{\alpha\beta}
\end{align}
and 
\begin{equation}
Q^\mu=p_\nu T^{\mu\nu},\qquad T=\delta_{\mu\nu}T^{\mu\nu}, \qquad Q= p_\nu p_\mu T^{\mu\nu}.
\end{equation}

One can consider the decomposition of the $\braket{TOOO}$ correlation function as
\begin{align}
\braket{T^{\mu_1\nu_1}(\mathbf{\bar{p}}_1)O(\mathbf{p}_2)O(\mathbf{p}_3)O(\mathbf{p}_4)}&=\braket{t^{\mu_1\nu_1}(\mathbf{\bar{p}}_1)O(\mathbf{p}_2)O(\mathbf{p}_3)O(\mathbf{p}_4)}+\braket{t_{loc}^{\mu_1\nu_1}(\mathbf{\bar{p}}_1)O(\mathbf{p}_2)O(\mathbf{p}_3)O(\mathbf{p}_4)}\label{decomp}
\end{align}
where in bold we refer to vectors in the Euclidean $\mathbb{R}^d$ space, and we are considering $\bar{p}_1^\mu=-p_2^\mu-p_3^\mu-p_4^\mu$ from momentum conservation. \\
The first term on the right hand side of \eqref{decomp} is the $tt$ part of the correlation function, and the second represents the local ($loc$) part. The method consists in expanding the $tt$ sector into a minimal number of form factors, fixed by the symmetry of the correlator \cite{Bzowski:2013sza}. 
In our case the $tt$ and local parts take the form
\begin{align}
\braket{t^{\mu_1\nu_1}(\mathbf{\bar{p}}_1)O(\mathbf{p}_2)O(\mathbf{p}_3)O(\mathbf{p}_4)}&=\Pi^{\mu_1\nu_1}_{\alpha_1\beta_1}(\mathbf{\bar{p}}_1)\left[A\,p_2^{\alpha_1}p_2^{\beta_1}+A(p_2\leftrightarrow p_3)\,p_3^{\alpha_1}p_3^{\beta_1}+A(p_2\leftrightarrow p_4)\,p_4^{\alpha_1}p_4^{\beta_1}\right]\label{transvtrace}\\
\braket{t_{loc}^{\mu_1\nu_1}(\mathbf{\bar{p}}_1)O(\mathbf{p}_2)O(\mathbf{p}_3)O(\mathbf{p}_4)}&=\Sigma^{\mu_1\nu_1}_{\alpha_1\beta_1}(\mathbf{\bar{p}}_1)\braket{T^{\alpha_1\beta_1}(\mathbf{\bar{p}}_1)O(\mathbf{p}_2)O(\mathbf{p}_3)O(\mathbf{p}_4)}\label{local}.
\end{align}
From these expressions, one can observe that the local term is constrained by the conservation WI's \eqref{ConsWI}, which project on 3-point functions of the form $OOO$, as we have discussed in the previous section. Using \eqref{Sigma} and \eqref{ConsWI} in \eqref{local}, one can explicitly write the local term in the form
\begin{align}
\braket{t_{loc}^{\mu_1\nu_1}(\mathbf{\bar{p}}_1)O(\mathbf{p}_2)O(\mathbf{p}_3)O(\mathbf{p}_4)}&=\frac{2}{\bar{p}_1^2}\Big[-\bar{p}_1^{(\mu_1}p_2^{\nu_1)}\braket{O(\mathbf{p}_3+\mathbf{p}_4)O(\mathbf{p}_3)O(\mathbf{p}_4)}\notag\\
&\hspace{-4cm}-\bar{p}_1^{(\mu_1}p_3^{\nu_1)}\braket{O(\mathbf{p}_2+\mathbf{p}_4)O(\mathbf{p}_2)O(\mathbf{p}_4)}-\bar{p}_1^{(\mu_1}p_4^{\nu_1)}\braket{O(\mathbf{p}_2+\mathbf{p}_3)O(\mathbf{p}_2)O(\mathbf{p}_3)}\Big]+\frac{(d-\Delta)}{d-1}\pi^{\mu_1\nu_1}(\mathbf{\bar{p}}_1)\notag\\
&\hspace{-4.7cm} \times\Big[\braket{O(\mathbf{p}_2+\mathbf{p}_4)O(\mathbf{p}_2)O(\mathbf{p}_4)}+\braket{O(\mathbf{p}_2+\mathbf{p}_3)O(\mathbf{p}_2)O(\mathbf{p}_3)}+\braket{O(\mathbf{p}_3+\mathbf{p}_4)O(\mathbf{p}_3)O(\mathbf{p}_4)}\Big]\notag\\
&\hspace{-4cm}-\frac{1}{(d-1)}\left(\delta^{\mu_1\nu_1}+(d-2)\frac{\bar{p}_1^{\mu_1}\bar{p}_1^{\nu_1}}{\bar{p}_1^2}\right)\left[-\frac{\mathbf{\bar{p}}_1\cdot\mathbf{p}_2}{\bar{p}_1^2}\braket{O(\mathbf{p}_3+\mathbf{p}_4)O(\mathbf{p}_3)O(\mathbf{p}_4)}
\right.\notag\\
&\hspace{-3cm}\left.-\frac{\mathbf{\bar{p}}_1\cdot\mathbf{p}_3}{\bar{p}_1^2}\braket{O(\mathbf{p}_2+\mathbf{p}_4)O(\mathbf{p}_2)O(\mathbf{p}_4)}-\frac{\mathbf{\bar{p}}_1\cdot\mathbf{p}_4}{\bar{p}_1^2}\braket{O(\mathbf{p}_2+\mathbf{p}_3)O(\mathbf{p}_2)O(\mathbf{p}_3)}\right].
\end{align}
The scalar 3-point function appearing on the right hand side is exactly known. In this way, the task of finding the structure of the entire $\braket{TOOO}$ has been reduced to the identification of only its $tt$ part. In particular, as we are going to show, all the WI's will constrain a single form factor.\\
The parameterization of this form factor (A), eventually, can be chosen according to 
the type of amplitude that one intends to consider, in order to facilitate the analysis. \\
For instance, in the case in which one in interested in a comparison between the conformal prediction and a free field theory realization  - for example in a 1 (graviton) $\to$ three (scalars) process - then it is convenient to adopt the parameterization $A\equiv A(p_2,p_3,p_4,s,t,u)$  and derive the equations 
using such variables. This choice is the one which respects the symmetries of the process, since the three scalars can be treated equally, and it allows to discuss more easily its asymptotic behaviour. Notice that in this case, momentum $p_1$ is treated as a dependent one $(\overline{p}_1)$ and needs to be differentiated implicitly in the corresponding equations. 
 
\section{Conformal Ward Identities in the $1\to 3$ formulation }
Using the $1\to 3$ symmetric formulation and the parameterization presented in \eqref{local}, the $A$ form factor exhibits the following symmetries 
\begin{align}
A(p_3\leftrightarrow p_4)&=A(p_2,p_3,p_4,s,t,u)\\
A(p_2\to p_4\to p_3\to p_2)&=A(p_2\leftrightarrow p_4)\\
A(p_2\to p_3\to p_4\to p_2)&=A(p_2\leftrightarrow p_3),
\end{align}
which can be written in the form
\begin{align}
A(p_2,p_4,p_3,s,u,t)&=A(p_2,p_3,p_4,s,t,u)\\
A(p_4,p_2,p_3,t,u,s)&=A(p_4,p_3,p_2,t,s,u)\\
A(p_3,p_4,p_2,u,s,t)&=A(p_3,p_2,p_4,u,t,s).
\end{align}

In order to extract some information on A$(p_2,p_3,p_4,s,t,u)$, we turn to the dilatation and the special conformal WI's which it has to satisfy. In the case of scalars of equal scaling $\Delta$ these take simplified forms respect to \eqref{dd1} and \eqref{SCWTJJ}

\begin{align}
0&=\left[(3\Delta-2d)-\sum_{j=2}^{4}\,p_j^\mu\,\frac{\partial}{\partial p_j^\mu}
\right]\,\braket{T^{\mu_1\nu_1}(\mathbf{\bar{p}}_1)O(\mathbf{p}_2)O(\mathbf{p}_3)O(\mathbf{p}_4)},\\
0&=\sum_{j=2}^4\left[2(\Delta-d)\frac{\partial}{\partial p_{j\,\kappa}}-2p_j^\alpha\,\frac{\partial^2}{\partial p_j^\alpha\,\partial\,p_{j\,\kappa}}+p_j^\kappa\,\frac{\partial^2}{\partial\,p_j^\alpha\,\partial\,p_{j\,\alpha}}\right]\,\braket{T^{\mu_1\nu_1}(\mathbf{\bar{p}}_1)O(\mathbf{p}_2)O(\mathbf{p}_3)O(\mathbf{p}_4)},
\end{align}
where, as already mentioned, the momentum $p_1^\mu$ is taken as the dependent one. \\
As discussed in \cite{Coriano:2018bbe}, one of the external coordinates of the correlator can be set to vanish by translational symmetry, and its corresponding momentum, after Fourier transform, has to be taken as dependent on the other. For instance, in this case,  for convenience, we have chosen the coordinate of the stress-energy tensor to vanish ($x_1=0$), and taken its momentum as the dependent one ($p_1\to \overline{p}_1$). This implies that the spin part of the special conformal transformation will not act on the stress-energy tensor, and the action of this generator is  reduced to a pure scalar.\\
The differentiation is performed only respect to the independent momenta, using the chain rule while differentiating $\overline{p}_1$. This choice is optimal if we intend to derive symmetric equations for the $TOOO$, in which we treat the three scalar operators equally, as is the case if we intend to investigate this correlator in a $1\to 3$ kinematical configuration. In section \ref{asymtreat} we will reverse this choice, by taking one of the scalar momenta ($p_4$) as the dependent one, which is equivalent to choosing $x_4=0$ in coordinate space. In this second case the special conformal generator will act with its spin part on the indices of the stress-energy tensor as well, being the momentum $p_1$ one of the independent momenta.
\\
The procedure that we will apply in this case follows quite closely the approach implemented for 3-point functions, developed in \cite{Bzowski:2013sza}. Both equations are projected onto the transverse traceless sector using the $\Pi$ projector, whose action is endomorphic on this sector \cite{Bzowski:2013sza}. A more detailed discussion of this point can be found in \cite{Coriano:2018bbe}. \\
Henceforth, by applying $\Pi^{\rho_1\sigma_1}_{\mu_1\nu_1}(\mathbf{\bar{p}}_1)$ on the left of the dilatation and special conformal generators, we find 
\begin{equation}
\Pi^{\rho_1\sigma_1}_{\mu_1\nu_1}(\mathbf{\bar{p}}_1)\,\hat{D}\,\braket{t^{\mu_1\nu_1}(\mathbf{\bar{p}}_1)O(\mathbf{p}_2)O(\mathbf{p}_3)O(\mathbf{p}_4)}=0\label{Kdil}
\end{equation}
for the dilatation WI, and
\begin{align}
0&=\Pi^{\rho_1\sigma_1}_{\mu_1\nu_1}(\mathbf{\bar{p}}_1)\,\mathcal{K}^\kappa\,\left[\braket{t^{\mu_1\nu_1}(\mathbf{\bar{p}}_1)O(\mathbf{p}_2)O(\mathbf{p}_3)O(\mathbf{p}_4)}+\braket{t_{loc}^{\mu_1\nu_1}(\mathbf{\bar{p}}_1)O(\mathbf{p}_2)O(\mathbf{p}_3)O(\mathbf{p}_4)}\right]\notag\\
&=\Pi^{\rho_1\sigma_1}_{\mu_1\nu_1}(\mathbf{\bar{p}}_1)\,\mathcal{K}^\kappa\,\braket{t^{\mu_1\nu_1}(\mathbf{\bar{p}}_1)O(\mathbf{p}_2)O(\mathbf{p}_3)O(\mathbf{p}_4)}+\Pi^{\rho_1\sigma_1}_{\mu_1\nu_1}\left[\frac{4d}{\bar{p}_1^2}\bar{p}_{1\,\beta}\delta^{\mu_1\kappa}\braket{T^{\nu_1\beta}(\mathbf{\bar{p}}_1)O(\mathbf{p}_2)O(\mathbf{p}_3)O(\mathbf{p}_4)}\right]\label{prim}
\end{align}
for the conformal WI,
where we have used the relation
\begin{equation}
\Pi^{\rho_1\sigma_1}_{\mu_1\nu_1}\,{\cal K}^\kappa\,\braket{t_{loc}^{\mu_1\nu_1}(\mathbf{\bar{p}}_1)O(\mathbf{p}_2)O(\mathbf{p}_3)O(\mathbf{p}_4)}=\Pi^{\rho_1\sigma_1}_{\mu_1\nu_1}\left[\frac{4d}{\bar{p}_1^2}\,\delta^{\mu_1\kappa}\,\bar{p}_{1\,\beta}\,\braket{T^{\nu_1\beta}(\mathbf{\bar{p}}_1)O(\mathbf{p}_2)O(\mathbf{p}_3)O(\mathbf{p}_4)}\right].
\end{equation}
 The first term in \eqref{prim} can be explicitly written as
\begin{align}
\Pi^{\rho_1\sigma_1}_{\mu_1\nu_1}(\mathbf{\bar{p}}_1)\,\mathcal{K}^\kappa\,\braket{t^{\mu_1\nu_1}(\mathbf{\bar{p}}_1)O(\mathbf{p}_2)O(\mathbf{p}_3)O(\mathbf{p}_4)}&=\notag\\
&\hspace{-7.3cm}=\Pi^{\rho_1\sigma_1}_{\mu_1\nu_1}(\mathbf{\bar{p}}_1)\Bigg\{p_2^\kappa\Bigg[C_{11}\,p_2^{\mu_1} p_2^{\nu_1}+C_{12}\,p_3^{\mu_1} p_3^{\nu_1}+C_{13}\,p_4^{\mu_1} p_4^{\nu_1}\Bigg]+p_3^\kappa\Bigg[C_{21}\,p_2^{\mu_1} p_2^{\nu_1}+C_{22}\,p_3^{\mu_1} p_3^{\nu_1}+C_{23}\,p_4^{\mu_1} p_4^{\nu_1}\Bigg]\notag\\
&\hspace{-4cm}+p_4^\kappa\Bigg[C_{31}\,p_2^{\mu_1} p_2^{\nu_1}+C_{32}\,p_3^{\mu_1} p_3^{\nu_1}+C_{33}\,p_4^{\mu_1} p_4^{\nu_1}\Bigg]+\delta^{\mu_1\kappa}\Bigg[C_{41}\,p_2^{\nu_1}+C_{42}\,p_3^{\nu_1}+C_{43}\,p_4^{\nu_1}\Bigg]\Bigg\}\label{Ktt}
\end{align}
where we have used the chain rules
\begin{align}
\frac{\partial}{\partial p_2^\mu}&=\frac{p_2^\mu}{p_2}\frac{\partial}{\partial p_2}+\frac{p_2^{\mu}+p_3^{\mu}}{t}\frac{\partial}{\partial t}+\frac{p_2^\mu+p_4^\mu}{u}\frac{\partial}{\partial u}\\
\frac{\partial}{\partial p_3^\mu}&=\frac{p_3^\mu}{p_3}\frac{\partial}{\partial p_3}+\frac{p_3^{\mu}+p_4^{\mu}}{s}\frac{\partial}{\partial s}+\frac{p_2^\mu+p_3^\mu}{t}\frac{\partial}{\partial t}\\
\frac{\partial}{\partial p_4^\mu}&=\frac{p_4^\mu}{p_4}\frac{\partial}{\partial p_4}+\frac{p_2^{\mu}+p_4^{\mu}}{s}\frac{\partial}{\partial s}+\frac{p_2^\mu+p_4^\mu}{u}\frac{\partial}{\partial u}
\end{align}
in order to write the covariant derivatives in terms of scalar derivatives involving the invariants parameterizing $A$. The coefficients $C_{ij}$ in \eqref{Ktt} are linear combinations of differential operators acting on $A$.
The dilatation WI \eqref{Kdil} can be written in scalar form as
\begin{align}
0&=\Pi^{\rho_1\sigma_1}_{\mu_1\nu_1}(\mathbf{\bar{p}}_1)\,\hat{D}\,\braket{t^{\mu_1\nu_1}(\mathbf{\bar{p}}_1)O(\mathbf{p}_2)O(\mathbf{p}_3)O(\mathbf{p}_4)}\notag\\
&=\Pi^{\rho_1\sigma_1}_{\mu_1\nu_1}(\mathbf{\bar{p}}_1)\Bigg\{D_1\,p_2^{\mu_1} p_2^{\nu_1}+D_2\,p_3^{\mu_1} p_3^{\nu_1}+D_3\,p_4^{\mu_1} p_4^{\nu_1}\Bigg]\Bigg\}
\end{align}
where $D_i$ are terms involving scalar derivatives acting on the form factor $A(p_2,p_3,p_4,s,t,u)$. The previous equation is satisfied if all the $D_i$ vanish independently, giving a dilatation constraint on $A(p_2,p_3,p_4,s,t,u)$ of the form
\begin{align}
D_1=0\,\implies \left[\sum_{j=2}^4\,p_j\frac{\partial}{\partial p_j}+s\frac{\partial}{\partial s}+t\frac{\partial}{\partial t}+u\frac{\partial}{\partial u}\right]A(p_i,s,t,u)=(\D_t-3d-2)\,A(p_i,s,t,u),
\end{align} 
where $\D_t=\sum_{j=1}^4\D_j=d+3\Delta$, since $\D_1=d$ for the stress-energy tensor, and we have set  $\Delta_2=\Delta_3=\Delta_4=\Delta$. From the other conditions $D_i=0$, with $i=2,3$, we generate the same constraint as from $D_1$, modulo some permutations 
involving $(p_2\leftrightarrow p_3)$ and $(p_2\leftrightarrow p_4)$ respectively. 

\subsection{Primary Conformal Ward Identities}
From the expressions of \eqref{prim} and \eqref{Ktt}, after some lengthy algebraic manipulations, we derive the primary constraints as
\begin{align}
&C_{11}=0,&&C_{12}=0,&&C_{13}=0,\notag\\
&C_{21}=0,&&C_{22}=0,&&C_{23}=0,\notag\\
&C_{31}=0,&&C_{32}=0,&&C_{33}=0,
\end{align}
which are explicitly given in \appref{AppendixB}.

\begin{figure}[h]
    \centering
\raisebox{-0.5\height}{\includegraphics[scale=0.8]{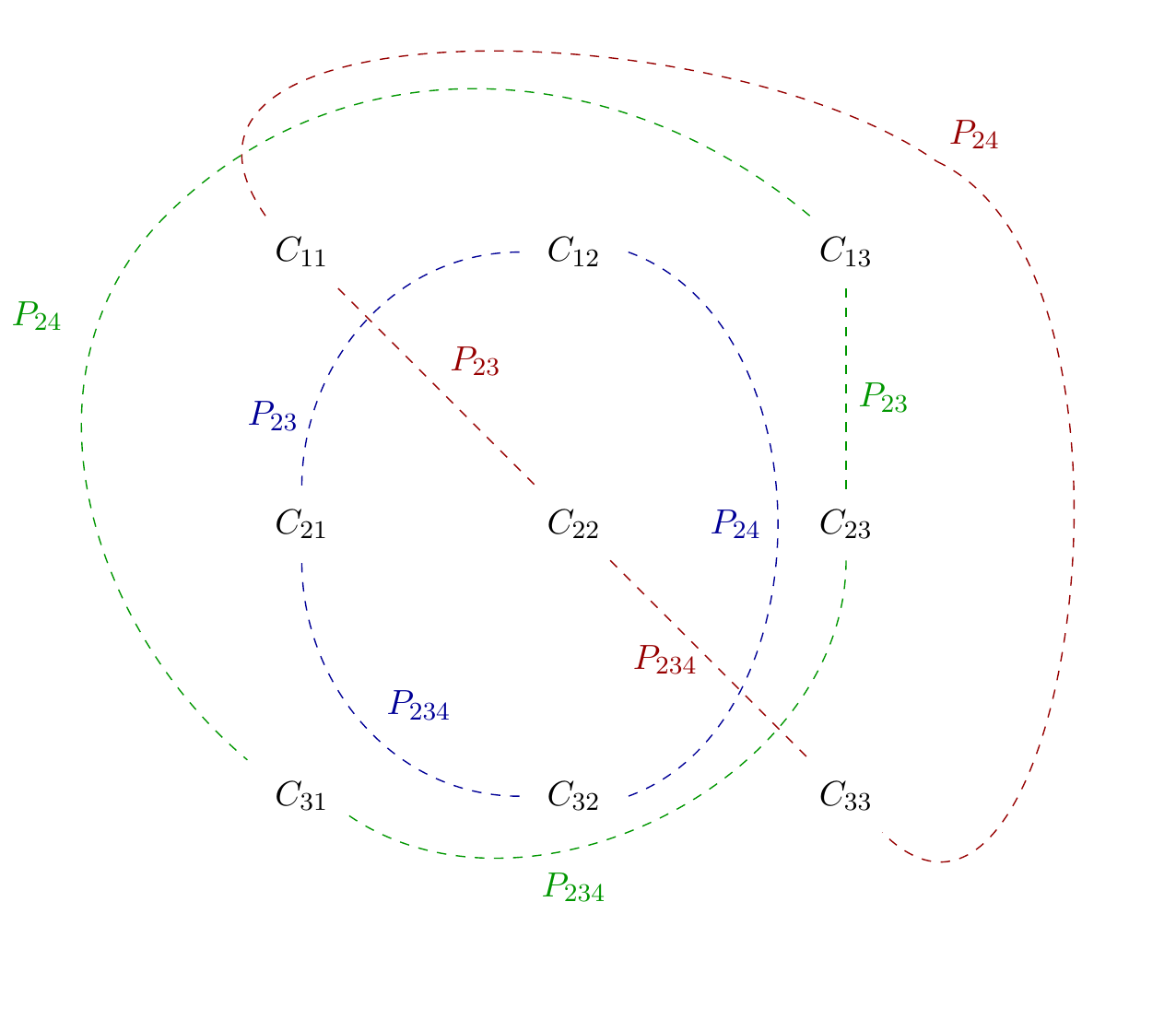}}
    \caption{Orbits of the primary CWI's of the $TOOO$ under $P_{23}$ and $P_{24}$}
    \label{tikzz}
\end{figure}

One can easily reorganize these equations by introducing the operators
\begin{align}
\bar{K}(p_2,p_3,p_4,s,t,u)&\equiv K_2+\frac{p_3^2-p_4^2}{s\,t}\frac{\partial}{\partial s \partial t}-\frac{p_3^2-p_4^2}{s\,u}\frac{\partial}{\partial s \partial u}+\frac{1}{t}\frac{\partial}{\partial t}\left(p_2\frac{\partial}{\partial p_2}+p_3\frac{\partial}{\partial p_3}-p_4\frac{\partial}{\partial p_4}\right)\notag\\
&\hspace{-3cm}+(d-\Delta)\left(\frac{1}{t}\frac{\partial}{\partial t}+\frac{1}{u}\frac{\partial}{\partial u}\right)+\frac{1}{u}\frac{\partial}{\partial u}\left(p_2\frac{\partial}{\partial p_2}-p_3\frac{\partial}{\partial p_3}+p_4\frac{\partial}{\partial p_4}\right)+\frac{2p_2^2+p_3^2+p_4^2-s^2-t^2-u^2}{t\,u}\frac{\partial}{\partial t\partial u}
\end{align}
and
\begin{equation}
L(s,t)\equiv\frac{2}{s}\frac{\partial}{\partial s}-\frac{2}{t}\frac{\partial}{\partial t}
\end{equation}
with
\begin{align}
L(s,t)=-L(t,s),
\end{align}
obtaining
\begin{align}
C_{11}&=\bar{K}(p_2,p_3,p_4,s,t,u)\,A(p_2,p_3,p_4,s,t,u)\notag\\
C_{12}&=\bar{K}(p_2,p_3,p_4,s,t,u)\,A(p_3,p_2,p_4,u,t,s)+L(t,u)\bigg(A(p_2,p_3,p_4,s,t,u)+A(p_3,p_2,p_4,u,t,s)\bigg)\notag\\
C_{13}&=\bar{K}(p_2,p_3,p_4,s,t,u)\,A(p_4,p_3,p_2,t,s,u)-L(t,u)\bigg(A(p_2,p_3,p_4,s,t,u)+A(p_4,p_3,p_2,t,s,u)\bigg)\notag\\[2ex]
C_{21}&=\bar{K}(p_3,p_2,p_4,u,t,s)\,A(p_2,p_3,p_4,s,t,u)-L(s,t)\bigg(A(p_2,p_3,p_4,s,t,u)+A(p_3,p_2,p_4,u,t,s)\bigg)\notag\\
C_{22}&=\bar{K}(p_3,p_2,p_4,u,t,s)\,A(p_3,p_2,p_4,u,t,s)\notag\\
C_{23}&=\bar{K}(p_3,p_2,p_4,u,t,s)\,A(p_4,p_3,p_2,t,s,u)+L(s,t)\bigg(A(p_4,p_3,p_2,t,s,u)+A(p_3,p_2,p_4,u,t,s)\bigg)\notag
\end{align}
\begin{align}
C_{31}&=\bar{K}(p_4,p_3,p_2,t,s,u)\,A(p_2,p_3,p_4,s,t,u)-L(s,u)\bigg(A(p_2,p_3,p_4,s,t,u)+A(p_4,p_3,p_2,t,s,u)\bigg)\notag\\
C_{32}&=\bar{K}(p_4,p_3,p_2,u,t,s)\,A(p_3,p_2,p_4,u,t,s)+L(s,u)\bigg(A(p_3,p_2,p_4,u,t,s)+A(p_4,p_3,p_2,t,s,u)\bigg)\notag\\
C_{33}&=\bar{K}(p_4,p_3,p_2,u,t,s)\,A(p_4,p_3,p_2,t,s,u).\notag
\end{align}
We illustrate in \eqref{tikzz} pictorially the action of the permutation operators $P_{ij}$, acting on the two momenta $p_i^\mu$ and $p_j^\mu$, on the various $C_{ij}$ presented above. The functional dependence of the form factor $A(p_2,p_3, p_4,s,t,u)$ will vary accordingly. \\
The orbits connect the various coefficients $C_{ij}$ which can be reached by the action of the various permutations. We start with $P_{23}$, $P_{24}$ and their product $P_{234}$.
 The orbits describe equivalent equations and we are allowed to choose any of the equations labelled by coefficients $C_{ij}$ belonging to separate orbits. Since there are three independent orbits under this subgroup, we start by selecting only three primary conformal WI's which are not related by the action of such permutations

\begin{figure}[h]
    \centering
\raisebox{-0.5\height}{\includegraphics[scale=0.8]{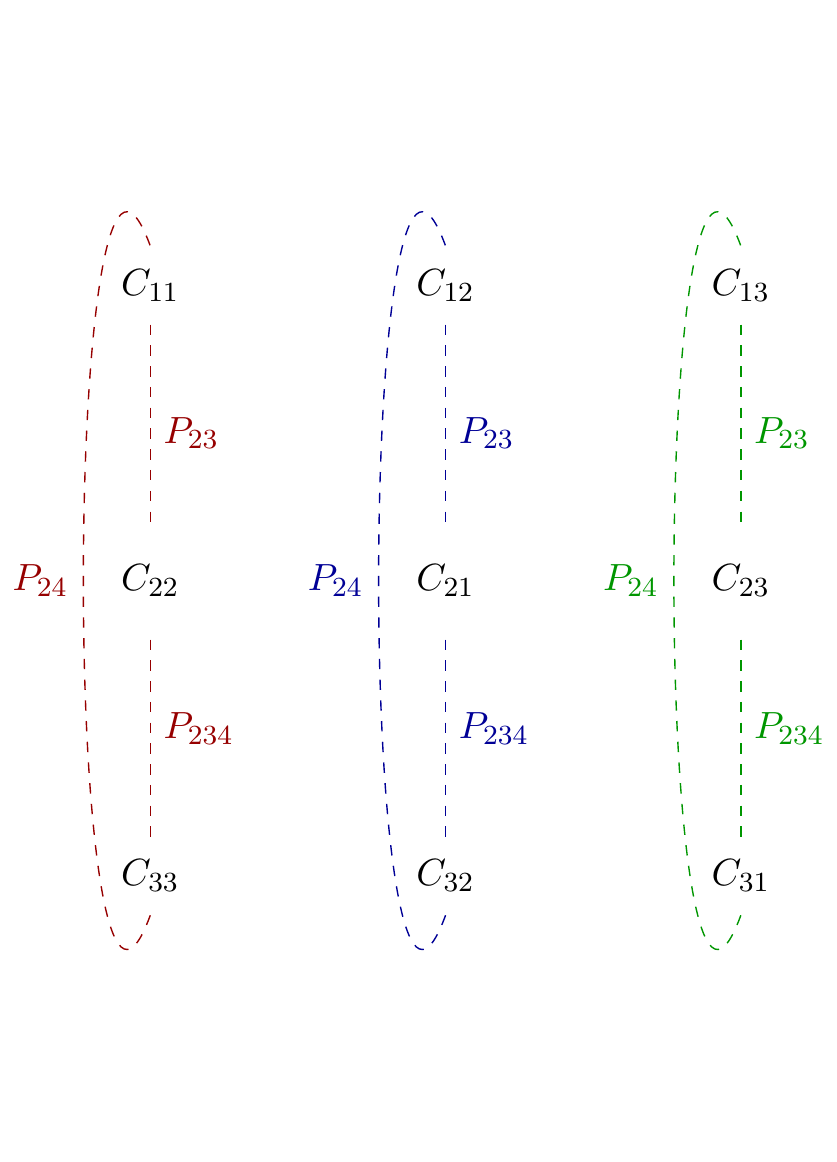}}
    \caption{Equivalent pictorial representation of the orbits as in Fig. \ref{tikzz}. }
    \label{}
\end{figure}
The equations that we are going to choose are
\begin{align}
C_{11}&=\bar{K}(p_2,p_3,p_4,s,t,u)\,A(p_2,p_3,p_4,s,t,u)\notag\\
C_{12}&=\bar{K}(p_2,p_3,p_4,s,t,u)\,A(p_3,p_2,p_4,u,t,s)+L(t,u)\bigg(A(p_2,p_3,p_4,s,t,u)+A(p_3,p_2,p_4,u,t,s)\bigg)\notag\\
C_{13}&=\bar{K}(p_2,p_3,p_4,s,t,u)\,A(p_4,p_3,p_2,t,s,u)-L(t,u)\bigg(A(p_2,p_3,p_4,s,t,u)+A(p_4,p_3,p_2,t,s,u)\bigg).
\end{align}
At this stage we include $P_{34}$, under whose action $C_{11}$ is mapped to itself, while $C_{12}\leftrightarrow C_{13}$. The mapping is illustrated below showing that the independent equations are only two.

\begin{figure}[h]
    \centering
\raisebox{-0.5\height}{\includegraphics[scale=1.2]{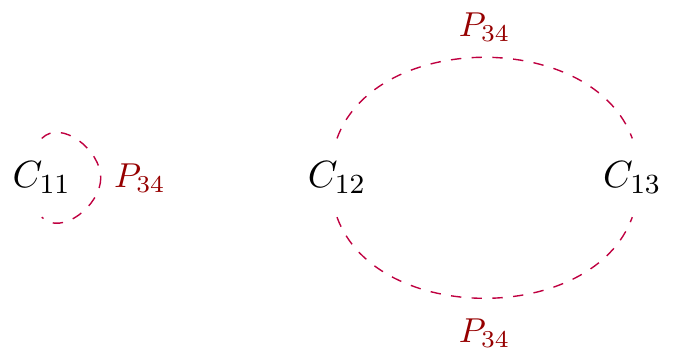}}
    \caption{Orbits of the CWI's for the $TOOO$ under $P_{34}$.}
    \label{three}
\end{figure}
 We take $C_{11}$ and $C_{12}$ as the independent ones, and all the other equations are obtained by acting on these two with a generic permutation of $(p_2,p_3,p_4)$. Therefore we have to solve only the two equations
\begin{equation}
\left\{
\begin{split}
0=&\bar{K}(p_2,p_3,p_4,s,t,u)\,A(p_2,p_3,p_4,s,t,u)\\
0=&\bar{K}(p_2,p_3,p_4,s,t,u)\,A(p_3,p_2,p_4,u,t,s)+L(t,u)\bigg(A(p_2,p_3,p_4,s,t,u)+A(p_3,p_2,p_4,u,t,s)\bigg)
\end{split}\right.
\end{equation}
as representatives of the set of the CWI, after taking into account all the symmetry properties of $A$. They can equivalently be set into the form
\begin{equation}
\left\{
\begin{split}
0&=\bar{K}(p_2,p_3,p_4,s,t,u)\,A(p_2,p_3,p_4,s,t,u)\\
0&=\bar{K}(p_3,p_2,p_4,u,t,s)\,A(p_2,p_3,p_4,s,t,u)-L(s,t)\bigg(A(p_2,p_3,p_4,s,t,u)+A(p_3,p_2,p_4,u,t,s)\bigg)
\end{split}\right.
\end{equation}
using the symmetry of the correlator. 
\subsection{Secondary Conformal Ward Identities}
\begin{figure}[h]
    \centering
\raisebox{-0.5\height}{\includegraphics[scale=1.2]{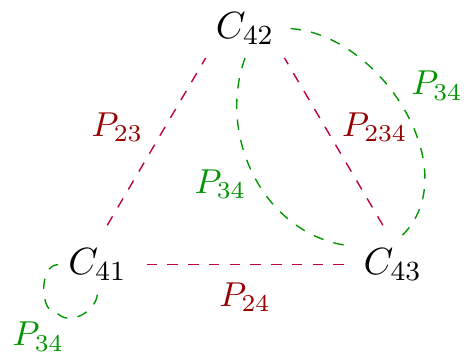}}
    \caption{Orbits of the secondary CWI's under permutations.}
    \label{four}
\end{figure}

The secondary CWI's for the correlator are first order differential equations derived from the coefficients $C_{4i}$, $i=1,\,2,\,3$ in \eqref{Ktt} together with Eq. \eqref{prim}. Such coefficients take the forms
\begin{align}
C_{41}&=\left[\frac{2(p_3^2-p_2^2-t^2)}{t}\frac{\partial}{\partial t}+\frac{2(p_4^2-p_2^2-u^2)}{u}\frac{\partial}{\partial u}-4\,p_2\,\frac{\partial}{\partial p_2}+\frac{2}{\bar{p}_1^2}\bigg(\frac{d(d-2)(p_2^2-s^2)-2s^2}{(d-1)}\bigg)\right.\notag\\
&\hspace{1cm}\left.+\frac{4\Delta(d-1)-2d^2-3(d-2)}{(d-1)}-\frac{(d-2)}{(d-1)}\frac{\big(p_2^2-s^2\big)^2}{\bar{p}_1^4}\right]\,A(p_2,p_3,p_4,s,t,u)
\notag\\
&-\left[\frac{2}{\bar{p}_1^2}\bigg(\frac{d(p_3^2-u^2)+2u^2}{(d-1)}\bigg)+\frac{(d-2)}{(d-1)}+\frac{(d-2)}{(d-1)}\frac{(p_3^2-u^2)^2}{\bar{p}_1^4}\right]\,A(p_3,p_2,p_4,u,t,s)\notag\\
&-\left[\frac{2}{\bar{p}_1^2}\bigg(\frac{d(p_4^2-t^2)+2t^2}{(d-1)}\bigg)+\frac{(d-2)}{(d-1)}+\frac{(d-2)}{(d-1)}\frac{(p_4^2-t^2)^2}{\bar{p}_1^4}\right]\,A(p_4,p_3,p_2,t,s,u)\,,
\end{align}

\begin{align}
C_{42}&=\left[\frac{2(p_2^2-p_3^2-t^2)}{t}\frac{\partial}{\partial t}+\frac{2(p_4^2-p_3^2-s^2)}{s}\frac{\partial}{\partial s}-4\,p_3\,\frac{\partial}{\partial p_3}+\frac{2}{\bar{p}_1^2}\bigg(\frac{d(d-2)(p_3^2-u^2)-2u^2}{(d-1)}\bigg)\right.\notag\\
&\hspace{1cm}\left.+\frac{4\Delta(d-1)-2d^2-3(d-2)}{(d-1)}-\frac{(d-2)}{(d-1)}\frac{\big(p_3^2-u^2\big)^2}{\bar{p}_1^4}\right]\,A(p_3,p_2,p_4,u,t,s)
\notag\\
&-\left[\frac{2}{\bar{p}_1^2}\bigg(\frac{d(p_2^2-s^2)+2s^2}{(d-1)}\bigg)+\frac{(d-2)}{(d-1)}+\frac{(d-2)}{(d-1)}\frac{(p_2^2-s^2)^2}{\bar{p}_1^4}\right]\,A(p_2,p_3,p_4,s,t,u)\notag\\
&-\left[\frac{2}{\bar{p}_1^2}\bigg(\frac{d(p_4^2-t^2)+2t^2}{(d-1)}\bigg)+\frac{(d-2)}{(d-1)}+\frac{(d-2)}{(d-1)}\frac{(p_4^2-t^2)^2}{\bar{p}_1^4}\right]\,A(p_4,p_3,p_2,t,s,u)\notag\\
\end{align}
and 
\begin{align}
C_{43}&=\left[\frac{2(p_3^2-p_4^2-s^2)}{s}\frac{\partial}{\partial s}+\frac{2(p_2^2-p_4^2-u^2)}{u}\frac{\partial}{\partial u}-4\,p_4\,\frac{\partial}{\partial p_4}+\frac{2}{\bar{p}_1^2}\bigg(\frac{d(d-2)(p_4^2-t^2)-2t^2}{(d-1)}\bigg)\right.\notag\\
&\hspace{1cm}\left.+\frac{4\Delta(d-1)-2d^2-3(d-2)}{(d-1)}-\frac{(d-2)}{(d-1)}\frac{\big(p_4^2-t^2\big)^2}{\bar{p}_1^4}\right]\,A(p_4,p_3,p_2,t,s,u)
\notag\\
&-\left[\frac{2}{\bar{p}_1^2}\bigg(\frac{d(p_3^2-u^2)+2u^2}{(d-1)}\bigg)+\frac{(d-2)}{(d-1)}+\frac{(d-2)}{(d-1)}\frac{(p_3^2-u^2)^2}{\bar{p}_1^4}\right]\,A(p_3,p_2,p_4,u,t,s)\notag\\
&-\left[\frac{2}{\bar{p}_1^2}\bigg(\frac{d(p_2^2-s^2)+2s^2}{(d-1)}\bigg)+\frac{(d-2)}{(d-1)}+\frac{(d-2)}{(d-1)}\frac{(p_2^2-s^2)^2}{\bar{p}_1^4}\right]\,A(p_2,p_3,p_4,s,t,u)\,,
\end{align}
where here $p_1^2$ is treated as a dependent variable, that is: $\bar{p}_1^2=s^2+t^2+u^2-p_2^2-p_3^2-p_4^2$. 
The actions of the operators enforcing the momentum permutations and the orbits of the $C_{ij}$ are illustrated in 
Fig. \eqref{four}, where a given equation is connected by a link if there is a permutation of the momenta which relates it to a different one. \\
It is clear from the figure that each single vertex of the triangle is mapped into itself under a permutation 
acting on the opposite edge, showing that there is only one independent secondary CWI. In particular, we choose as the independent one $C_{41}$, which can be re-expressed in the form

\begin{align}
&L_1\,A(p_2,p_3,p_4,s,t,u)\notag\\
&\qquad-L_2\,A(p_3,p_2,p_4,u,t,s)-L_2(p_3\leftrightarrow p_4)\,A(p_4,p_3,p_2,t,s,u)=-\frac{4d}{\bar{p}_1^2}\braket{O(\mathbf{p}_3+\mathbf{p}_4)O(\mathbf{p}_3)O(\mathbf{p}_4)},\label{secondary}
\end{align}
where 
\begin{align}
L_1&=\left[\frac{2(p_3^2-p_2^2-t^2)}{t}\frac{\partial}{\partial t}+\frac{2(p_4^2-p_2^2-u^2)}{u}\frac{\partial}{\partial u}-4\,p_2\,\frac{\partial}{\partial p_2}+\frac{2}{\bar{p}_1^2}\bigg(\frac{d(d-2)(p_2^2-s^2)-2s^2}{(d-1)}\bigg)\right.\notag\\
&\hspace{1cm}\left.+\frac{4\Delta(d-1)-2d^2-3(d-2)}{(d-1)}-\frac{(d-2)}{(d-1)}\frac{\big(p_2^2-s^2\big)^2}{\bar{p}_1^4}\right]\\
L_2&=\left[\frac{2}{\bar{p}_1^2}\bigg(\frac{d(p_3^2-u^2)+2u^2}{(d-1)}\bigg)+\frac{(d-2)}{(d-1)}+\frac{(d-2)}{(d-1)}\frac{(p_3^2-u^2)^2}{\bar{p}_1^4}\right].
\end{align}
From \eqref{secondary}, one can check that the symmetry $p_3\leftrightarrow p_4$ is explicitly manifest.\\
In general, the role of the secondary WI's is to reduce the parameters of the solutions of the primary ones. For instance, in the case of 3-point functions, such solutions are determined from the primary equations modulo few constants, which are then fixed by the secondary ones. In that case, the right hand side of the secondary equations will be proportional to 2-point functions.\\
 The constraints on the primary solutions are obtained by taking special limits on the left hand side of the equations, in order to send two external coordinates into coalescence. This is obtained, for 3-point functions,  by taking two of the external invariant masses large and of unit ratio - $p_3^2/p_2^2\to 1$, for instance - which reduces the correlator to a 2-point function. \\
For 4-point functions this limit is far more involved, and we will be able to say little about it, the crucial point being that the primary solutions should contain arbitrary function(s), in this case a single function, as expected from the analysis in coordinate space, which are not identified in our formulation.\\
For this reason, we will try to discuss the asymptotic limit only of the primary solutions, where it is possible to underscore some specific behaviors of such solutions just by examining the structure of the equations. 

\section{The decomposition of the $TOOO$ in the $2\rightarrow2$ formulation}
\label{asymtreat}
In this section we will reconsider the  $TOOO$ correlator  $\langle T^{\mu_1 \nu_1}(\bold{p_1})O(\bold{p_2})O(\bold{p_3})O(\mathbf{\bar{p}_4})\rangle $ but with a different choice of the dependent momentum compared to the $1\to 3$ case,  which is suitable for the study of a $2\to 2$ process.\\
 We choose $p_4^{\mu}$ as the dependent momentum, $\bar{p_4}^{\mu}=-p_1^{\mu}-p_2^{\mu}-p_3^{\mu}$. Moreover, also the Mandelstam invariant $u^2$, will be taken as dependent variable  $\tilde{u}^2=-s^2-t^2+\sum p_i^2$.\\
We rewrite the decomposition \eqref{decomp} in the form

\begin{equation}\label{tensordecomp}
\langle T^{\mu_1 \nu_1}(\bold{p_1})O(\bold{p_2})O(\bold{p_3})O(\mathbf{\bar{p}_4})\rangle=\langle t^{\mu_1 \nu_1}(\bold{p_1})O(\bold{p_2})O(\bold{p_3})O(\mathbf{\bar{p}_4})\rangle +
\langle t^{\mu_1 \nu_1}_{\textrm{loc}}(\bold{p_1})O(\bold{p_2})O(\bold{p_3})O(\mathbf{\bar{p}_4})\rangle,
\end{equation} 
which is symmetric in $p_2^{\mu},p_3^{\mu},p_4^{\mu}$. Furthermore, we require  the parameterization in the $tt$ sector to be symmetric under the exchange of the indices of the stress-energy tensor $\mu_1 \leftrightarrow \nu_1$. 
The $tt$ component of the $TOOO$ can then be parameterized as
\begin{equation}
\langle t^{\mu_1 \nu_1}(\bold{p_1})O(\bold{p_2})O(\bold{p_3})O(\mathbf{\bar{p}_4})\rangle=\Pi^{\mu_1 \nu_1}_{\alpha_1 \beta_1}(\bold{p_1})X^{\alpha_1 \beta_1},
\end{equation}
where $X^{\alpha_1 \beta_1}$ is a general rank-2  tensor built out momenta and Kronecker's delta's. There are two equivalent decompositions of such $tt$ term, that we will present below, but only one of them allows to obtain simplified expressions of the primary and secondary CWI's, which will turn very useful for our analsysis.

\subsubsection{First decomposition}
First, we are going to derive the decomposition of the correlator by choosing as independent  momenta $p_2^{\mu},p_3^{\mu}$.  Terms that include $p_1^{\mu}$ will be eliminated by the transverse-traceless projector $\Pi^{\mu_1 \nu_1}_{\alpha_1 \beta_1}(\bold{p_1})$ and therefore will be omitted. We obtain the parameterization
\begin{equation}
X^{\alpha_1 \beta_1}=C^{\prime}(p_1,p_2,p_3,p_4,s,t)p_2^{\alpha_1}p_2^{\beta_1}+C^{\prime \prime}(p_1,p_2,p_3,p_4,s,t)p_3^{\alpha_1}p_3^{\beta_1}+C(p_1,p_2,p_3,p_4,s,t) p_2^{\alpha_1} p_3^{\beta_1},
\end{equation}
expressed in temrs of form factors $C, C^\prime, C^{\prime\prime}$.
Now, by imposing all the possible permutations (6 in total) of the momenta $p_2^{\mu},p_3^{\mu},p_4^{\mu}$, we derive the constraints
\begin{align*}
&C^{\prime}(p_1,p_2,p_3,p_4,s,t)=\frac{1}{2}\Big(C(p_1,p_2,p_3,p_4,s,t)+C(p_1,p_4,p_3,p_2,t,s)\Big),\\
&C^{\prime \prime}(p_1,p_2,p_4,p_3,s,\tilde{u})=\frac{1}{2}\Big(C(p_1,p_2,p_3,p_4,s,t)+C(p_1,p_2,p_4,p_3,s,\tilde{u})\Big).
\end{align*}
Finally, such $tt$ component takes the form
\begin{equation}\label{first}
\begin{split}
\langle t^{\mu_1 \nu_1}(\bold{p_1})O(\bold{p_2})O(\bold{p_3})O(\mathbf{\bar{p}_4})\rangle&=\Pi^{\mu_1 \nu_1}_{\alpha_1 \beta_1}\Bigg[\frac{1}{2}\Big(C(p_1,p_2,p_3,p_4,s,t)+C(p_1,p_4,p_3,p_2,t,s)\Big)p_2^{\alpha_1}p_2^{\beta_1}\\&
+\frac{1}{2}\Big(C(p_1,p_2,p_3,p_4,s,t)+C(p_1,p_2,p_4,p_3,s,\tilde{u})\Big)p_3^{\alpha_1}p_3^{\beta_1}+C(p_1,p_2,p_3,p_4,s,t) p_2^{\alpha_1} p_3^{\beta_1}\Bigg],
\end{split}
\end{equation}
 expressed in terms of a single form factor which exhibits the following symmetries
\begin{equation}
\begin{split}
\label{cc}
&C(p_1,p_2,p_3,p_4,s,t)=C(p_1,p_3,p_2,p_4,\tilde{u},t),\\
&C(p_1,p_2,p_4,p_3,s,\tilde{u})=C(p_1,p_4,p_2,p_3,\tilde{u},s)\\
&C(p_1,p_4,p_3,p_2,t,s)=C(p_1,p_3,p_4,p_2,\tilde{u},s).
\end{split}
\end{equation}

\subsubsection{Second decomposition}
The second decomposition is obtained by using all the available momenta. 
It takes the form
\begin{equation}
\langle t^{\mu_1 \nu_1}(\bold{p_1})O(\bold{p_2})O(\bold{p_3})O(\mathbf{\bar{p}_4})\rangle=\Pi^{\mu_1 \nu_1}_{\alpha_1 \beta_1}(\bold{p_1})\tilde{X}^{\alpha_1 \beta_1},
\end{equation}
where 
\begin{equation}
\tilde{X}^{\alpha_1 \beta_1}=F(p_1,p_2,p_3,p_4,s,t) p_2^{\alpha_1}p_3^{\beta_1}+F^{\prime}(p_1,p_2,p_3,p_4,s,t) p_2^{\alpha_1}p_4^{\beta_1}+F^{\prime \prime}(p_1,p_2,p_3,p_4,s,t) p_3^{\alpha_1}p_4^{\beta_1}.
\end{equation}
Taking into account all the possible permutations, we end up with the expression
\begin{equation}\label{second}
\tilde{X}^{\alpha_1 \beta_1}=F(p_1,p_4,p_3,p_2,t,s) p_3^{\alpha_1}p_4^{\beta_1}+F(p_1,p_2,p_4,p_3,s,\tilde{u}) p_2^{\alpha_1}p_4^{\beta_1}+F(p_1,p_2,p_3,p_4,s,t) p_2^{\alpha_1}p_3^{\beta_1}.
\end{equation}
Our form factor obeys the following symmetries:
\begin{equation}
\begin{split}
&F(p_1,p_3,p_4,p_2,\tilde{u},s)=F(p_1,p_4,p_3,p_2,t,s),\\
&F(p_1,p_4,p_2,p_3,\tilde{u},s)=F(p_1,p_2,p_4,p_3,s,\tilde{u}),\\
&F(p_1,p_3,p_2,p_4,\tilde{u},t)=F(p_1,p_2,p_3,p_4,s,t).
\end{split}
\end{equation}
Now, we can impose momentum conservation on the first two terms of \eqref{first}. Then comparing with \eqref{second} and using the symmetry properties of the previous form factor $C$ in \eqref{cc}, we obtain
\begin{align}
F(p_1,p_2,p_3,p_4,s,t)=-\frac{1}{2}\Big(C(p_1,p_2,p_3,p_4,s,t)+C(p_1,p_2,p_4,p_3,s,\tilde{u})\Big),\notag\\
F(p_1,p_2,p_4,p_3,s,\tilde{u})=-\frac{1}{2}\Big(C(p_1,p_2,p_3,p_4,s,t)+C(p_1,p_4,p_3,p_2,t,s)\Big),\notag\\
F(p_1,p_4,p_3,p_2,t,s)=-\frac{1}{2}\Big(C(p_1,p_2,p_3,p_4,s,t)+C(p_1,p_2,p_4,p_3,s,\tilde{u})\Big).
\label{relat}
\end{align}
The form factors $F$ and $C$ are related proving the equivalence between the two parameterizations. However $F$ is the one which generates CWI's of a simpler structure.

\subsection{Dilatation Ward Identity in the $2\rightarrow2$ formulation }
In this section we will proceed with the study of the dilatation WI. Using the form factor $F$, the full correlator is given by \eqref{tensordecomp} and the
exact parameterization of its $tt$ sector takes the form
\begin{equation}\label{TransverseFull}
\begin{split}
\langle t^{\mu_1 \nu_1}(\bold{p_1})O(\bold{p_2})O(\bold{p_3})O(\mathbf{\bar{p}_4})\rangle=&\Pi^{\mu_1 \nu_1}_{\alpha_1 \beta_1}(\bold{p_1})\Big(F(p_1,p_4,p_3,p_2,t,s) p_3^{\alpha_1}p_4^{\beta_1}+F(p_1,p_2,p_4,p_3,s,\tilde{u}) p_2^{\alpha_1}p_4^{\beta_1}\\&+F(p_1,p_2,p_3,p_4,s,t) p_2^{\alpha_1}p_3^{\beta_1}\Big),
\end{split}
\end{equation}
while the longitudinal sector is extracted by a contraction with the longitudinal projector
\begin{equation}\label{LocalFull}
\langle t^{\mu_1 \nu_1}_{\textrm{loc}}(\bold{p_1})O(\bold{p_2})O(\bold{p_3})O(\mathbf{\bar{p}_4})\rangle=\Sigma^{\mu_1 \nu_1}_{\alpha_1 \beta_1}(\bold{p_1})  \langle T^{\alpha_1 \beta_1}(\bold{p_1})O(\bold{p_2})O(\bold{p_3})O(\mathbf{\bar{p}_4})\rangle
\end{equation}
as in our previous analysis of the equations for the $1\to 3$.
We can express the CWI' s in terms of  6 invariants of the four-point function ($\sqrt{ p_i^2}=p_i$, $s=\sqrt{(\mathbf{p_1+p_2})^2},t=\sqrt{(\mathbf{p_2+p_3})^2}$) by using the chain rules
\begin{align}
& \frac{\partial}{\partial p_{1 \mu}}=\frac{p_1^{\mu}}{p_1}\frac{\partial}{\partial p_1}-\frac{\bar{p_4}^{\mu}}{p_4}\frac{\partial}{\partial p_4}+\frac{p_1^{\mu}+p_2^{\mu}}{s}\frac{\partial}{\partial s},\\
&\frac{\partial}{\partial p_{2 \mu}}=\frac{p_2^{\mu}}{p_2}\frac{\partial}{\partial p_1}-\frac{\bar{p_4}^{\mu}}{p_4}\frac{\partial}{\partial p_4}+\frac{p_1^{\mu}+p_2^{\mu}}{s}\frac{\partial}{\partial s}+\frac{p_2^{\mu}+p_3^{\mu}}{t}\frac{\partial}{\partial t},\\
&\frac{\partial}{\partial p_{3 \mu}}=\frac{p_3^{\mu}}{p_3}\frac{\partial}{\partial p_3}-\frac{\bar{p_4}^{\mu}}{p_4}\frac{\partial}{\partial p_4}+\frac{p_2^{\mu}+p_3^{\mu}}{t}\frac{\partial}{\partial t}.
\label{chain}
\end{align}
Applying the dilatation WI to \eqref{tensordecomp} we obtain
\begin{equation}\label{fulldilatation}
\big[(\Delta_{\textrm{t}}-3d)-\sum_{i=1}^{3} p_i^{\lambda} \frac{\partial}{\partial p_i^{\lambda}}\big]\langle T^{\mu_1 \nu_1}(\bold{p_1})O(\bold{p_2})O(\bold{p_3})O(\mathbf{\bar{p}_4})\rangle=0,
\end{equation}
which can be projected using the $tt$ projector $\Pi^{\rho \sigma}_{\mu_1 \nu_1}(\bold{p_1})$ obtaining  

\begin{equation}
\Pi^{\rho \sigma}_{\mu_1 \nu_1}(\bold{p_1})\hat{D}\langle t^{\mu_1 \nu_1}(\bold{p_1})O(\bold{p_2})O(\bold{p_3})O(\mathbf{\bar{p}_4})\rangle=0.
\end{equation}
Using  \eqref{TransverseFull} and differentiating by the chain rule \eqref{chain}, we finally obtain the equation \begin{equation}\label{DWI}
\left[(2+3d-\Delta_{\textrm{t}})+\sum_{i=1}^{4} p_i \frac{\partial }{\partial p_i}+s \frac{\partial }{\partial s}+t \frac{\partial  }{\partial t}\right]F(p_1,p_2,p_3,p_4,s,t)=0.
\end{equation}
The same equation holds also for $F(p_1,p_2,p_4,p_3,s,\tilde{u}),$ and $F(p_1,p_4,p_3,p_2,t,s)$.  The 2 in the first term of the sum $(2 + 3 d\ldots$), defines the tensorial dimension of the form factor, and counts the number of momenta with which it appears in the parameterization.

\subsection{Special CWI for the \texorpdfstring{ $\langle T^{\mu_1 \nu_1}(\bold{p_1})O(\bold{p_2})O(\bold{p_3})O(\mathbf{\bar{p}_4})\rangle $}{} }\label{scwilocal}
In this section we repeat the analysis of the $1\to 3$ case, with the new parameterization of the correlator that we have just derived, by selecting $p_4$ as the dependent momentum.
The action of the special conformal generator, as before, will take the form
\begin{equation}\label{CWI0}
\begin{split}
0&=\mathcal{K}^{\kappa}\braket{ T^{\mu_1 \nu_1}(\bold{p_1})O(\bold{p_2})O(\bold{p_3})O(\mathbf{\bar{p}_4})}\\[1.2ex]
&=\mathcal{K}^{\kappa}\langle t^{\mu_1 \nu_1}(\bold{p_1})O(\bold{p_2})O(\bold{p_3})O(\bar{\bold{p_4}})\rangle +\mathcal{K}^{\kappa}
\braket{ t^{\mu_1 \nu_1}_{\textrm{loc}}(\bold{p_1})O(\bold{p_2})O(\bold{p_3})O(\bar{\bold{p_4}})},
\end{split}
\end{equation}
We will focus now on the local part related to $t_{loc}$.\\
Using \eqref{LocalFull}  we now have to compute
\begin{equation}
\Pi^{\rho \sigma}_{\mu_1 \nu_1}(p_1)\mathcal{K}^{\kappa}\Sigma^{\mu_1 \nu_1}_{\alpha_1 \beta_1}  \langle  T^{\alpha_1 \beta_1}(\bold{p_1})O(\bold{p_2})O(\bold{p_3})O(\mathbf{\bar{p}_4})\rangle ,
\end{equation}
projected on the $tt$ sector. 
We split our results into the scalar and the spin part of $\mathcal{K}^{\kappa}$. Acting with the projection $\Pi$ we obtain
\begin{equation}
\begin{split}
&\Pi^{\rho \sigma}_{\mu_1 \nu_1}(p_1)\mathcal{K}^{\kappa}_{\mathrm{scalar}}  \langle  t^{\mu_1 \nu_1}_{\textrm{loc}}(\bold{p_1})O(\bold{p_2})O(\bold{p_3})O(\mathbf{\bar{p}_4})\rangle=\\
&=\Pi^{\rho \sigma}_{\mu_1 \nu_1}(p_1)\Bigg[ 4(\Delta_1-d+1)\frac{\delta^{\kappa \mu_1}\delta^{\nu_1}_{\alpha_1}p_{1,\beta_1}}{p_1^2}+4\frac{p_1^{\kappa}}{p_1^2}\delta^{\mu_1}_{\beta_1}\delta^{\nu_1}_{\alpha_1} \Bigg]\langle T^{\alpha_1 \beta_1}(\bold{p_1})O(\bold{p_2})O(\bold{p_3})O(\mathbf{\bar{p}_4})\rangle,
\end{split}
\end{equation}
and
\begin{equation}
\begin{split}
&\Pi^{\rho \sigma}_{\mu_1 \nu_1}(p_1)\mathcal{K}^{\kappa}_{\alpha,\mathrm{spin}}  \langle  t^{\mu_1 \nu_1}_{\textrm{loc}}(\bold{p_1})O(\bold{p_2})O(\bold{p_3})O(\mathbf{\bar{p}_4})\rangle=\\
&=\Pi^{\rho \sigma}_{\mu_1 \nu_1}(\textbf{p}_1)\Bigg[4(d-1)\frac{\delta^{\kappa \mu_1}\delta^{\nu_1}_{\alpha_1}p_{1,\beta_1}}{p_1^2}-4\frac{p_1^{\kappa}}{p_1^2}\delta^{\mu_1}_{\beta_1}\delta^{\nu_1}_{\alpha_1}\Bigg]\langle  T^{\alpha_1 \beta_1}(\bold{p_1})O(\bold{p_2})O(\bold{p_3})O(\mathbf{\bar{p}_4})\rangle. 
\end{split}
\end{equation}
In our results, we have ignored terms that include $p_1^{\mu_1},p_1^{\nu_1},\delta^{\mu_1 \nu_1}$, because we have the freedom to apply a transverse-traceless projector of the form $\Pi^{\rho \sigma}_{\mu_1 \nu_1}(\mathbf{p}_1)$ to \eqref{CWI0}, so these terms will vanish. Adding the scalar and the spin contributions (and using $\Delta_1=d$), we get
\begin{equation}\label{CWIlocal}
\Pi^{\rho \sigma}_{\mu_1 \nu_1}(p_1)\mathcal{K}^{\kappa}\langle  t^{\mu_1 \nu_1}_{\textrm{loc}}(\bold{p_1})O(\bold{p_2})O(\bold{p_3})O(\mathbf{\bar{p}_4})\rangle=\Pi^{\rho \sigma}_{\mu_1 \nu_1}(p_1)\left(\frac{4d}{p_1^2}\delta^{\kappa \mu_1}p_{1\,\alpha}\langle  T^{\alpha \nu_1}(\bold{p_1})O(\bold{p_2})O(\bold{p_3})O(\mathbf{\bar{p}_4})\rangle\right).
\end{equation}

Now, we will apply the $\mathcal{K}^{\k}$ operator on the $tt$ part, followed by contraction with the $\Pi$ projector. We obtain the tensor equation

\begin{equation}\label{fullform}
\begin{split}
&\Pi^{\rho \sigma}_{\mu_1 \nu_1}(p_1)\mathcal{K}^{\kappa} \langle t^{\mu_1 \nu_1}(\bold{p_1})O(\bold{p_2})O(\bold{p_3})O(\mathbf{\bar{p}_4})\rangle=\\
&\Pi^{\rho \sigma}_{\mu_1 \nu_1}(p_1)\Big[p_1^{\kappa}\big(\tilde{C}_{11}p_2^{\mu_1}p_3^{\nu_1}+\tilde{C}_{12}p_2^{\mu_1}p_4^{\nu_1}+\tilde{C}_{13}p_3^{\mu_1}p_4^{\mu_1}\big)+p_2^{\kappa}\big(\tilde{C}_{21}p_2^{\mu_1}p_3^{\nu_1}+\tilde{C}_{22}p_2^{\mu_1}p_4^{\nu_1}+\tilde{C}_{23}p_3^{\mu_1}p_4^{\mu_1}\big)\\
&+p_3^{\kappa}\big(\tilde{C}_{31}p_2^{\mu_1}p_3^{\nu_1}+\tilde{C}_{32}p_2^{\mu_1}p_4^{\nu_1}+\tilde{C}_{33}p_3^{\mu_1}p_4^{\mu_1}\big)+\delta^{\mu_1 \kappa}\big(\tilde{C}_{41}p_2^{\nu_1}+\tilde{C}_{42}p_3^{\nu_1}\big)+\delta^{\nu_1 \kappa}\big(\tilde{C}_{51}p_2^{\mu_1}+\tilde{C}_{52}p_3^{\mu_1}\big)\Big]
\end{split}
\end{equation}
which will allow us to extract the independent conformal constraints.
\subsection{Primary Conformal Ward Identities}\label{PCWI22}
\label{primd}
The factors $\tilde{C}_{1j},\tilde{C}_{2j},\tilde{C}_{3j}$ are second-order differential equations involving the form factor $F$ and its various permutations. We see from \eqref{CWI0} and \eqref{fullform} that the coefficients  of the four-momenta $p_1^{\kappa},p_2^{\kappa},p_3^{\kappa}$ are zero. This translates into the equations  
\begin{align}
&\tilde{C}_{11}=0,&&\tilde{C}_{12}=0,&&\tilde{C}_{13}=0,\notag\\
&\tilde{C}_{21}=0,&&\tilde{C}_{22}=0,&&\tilde{C}_{23}=0,\notag\\
&\tilde{C}_{31}=0,&&\tilde{C}_{32}=0,&&\tilde{C}_{33}=0.
\end{align}

These are the primary  CWI's that we have mentioned before. Below we present the explicit expressions involving the $F(p_1,p_2,p_3,p_4,s,t)$ form factor. The remaining ones, which are obtained just by permutations of the momenta, can be found in \appref{AppendixB}. We obtain

\begin{align}
\label{c11}
\tilde{C}_{11}=&\Bigg[\frac{\partial^2}{\partial p_4^2}+\frac{d-2\Delta+1}{p_4}\frac{\partial}{\partial p_4}-\frac{\partial^2}{\partial p_1^2}-\frac{1-d}{p_1}\frac{\partial}{\partial p_1}+\frac{1}{s}\frac{\partial}{\partial s}\left(p_4\frac{\partial}{\partial p_4}+p_3\frac{\partial}{\partial p_3}-p_1\frac{\partial}{\partial p_1}-p_2\frac{\partial}{\partial p_2}\right)\notag\\&
+\frac{d-\Delta}{s}\frac{\partial}{\partial s}
+\frac{p_3^2-p_2^2}{st}\frac{\partial^2}{\partial s \partial t}\Bigg]F(p_1,p_2,p_3,p_4,s,t)+\frac{2}{s}\frac{\partial F(p_1,p_4,p_3,p_2,t,s)}{\partial s}-\frac{2}{s}\frac{\partial F(p_1,p_2,p_4,p_3,s,\tilde{u})}{\partial s}
\end{align}
\begin{align}\label{c21}
\tilde{C}_{21}=&\Bigg[\frac{\partial^2 }{\partial p_4^2}+\frac{d-2\Delta+1}{p_4}\frac{\partial }{\partial p_4}-\frac{\partial^2 }{\partial p_2^2}-\frac{d-2\Delta+1}{p_2}\frac{\partial }{\partial p_2}+\frac{1}{s}\frac{\partial}{\partial s}\left(p_3\frac{\partial }{\partial p_3}+p_4\frac{\partial }{\partial p_4}-p_1\frac{\partial }{\partial p_1}-p_2\frac{\partial }{\partial p_2}\right)
\notag\\&
+\frac{\Delta-d-2}{t}\frac{\partial}{\partial t}+\frac{d-\Delta}{s}\frac{\partial}{\partial s}+\frac{1}{t}\frac{\partial}{\partial t}\left( p_1 \frac{\partial }{\partial p_1}+p_4\frac{\partial }{\partial p_4}-p_2\frac{\partial }{\partial p_2}-p_3\frac{\partial }{\partial p_3}\right)
\notag
\\&+\frac{p_4^2-p_2^2}{st}\frac{\partial^2}{\partial s \partial t}\Bigg]F(p_1,p_2,p_3,p_4,s,t)+\frac{2}{s}\frac{\partial F(p_1,p_4,p_3,p_2,t,s)}{\partial s}-\frac{2}{s}\frac{\partial F(p_1,p_2,p_4,p_3,s,\tilde{u})}{\partial s},\nonumber \\
\end{align}

and finally
\begin{align}\label{c31}
\tilde{C}_{31}=&\Bigg[\frac{\partial^2 }{\partial p_4^2}+\frac{d-2\Delta+1}{p_4}\frac{\partial }{\partial p_4}-\frac{\partial^2 }{\partial p_3^2}-\frac{d-2\Delta+1}{p_3}\frac{\partial }{\partial p_3}+\frac{2}{s}\frac{\partial}{\partial s}+\frac{p_1^2-p_2^2}{st}\frac{\partial^2}{\partial s \partial t}+\frac{\Delta-d-2}{t}\frac{\partial}{\partial t}\notag\\&+\frac{1}{t}\frac{\partial}{\partial t}\left(p_1\frac{\partial }{\partial p_1}+p_4\frac{\partial }{\partial p_4}-p_2\frac{\partial }{\partial p_2}-p_3\frac{\partial }{\partial p_3}\right)
\Bigg]F(p_1,p_2,p_3,p_4,s,t)+\frac{2}{s}\frac{\partial F(p_1,p_4,p_3,p_2,t,s)}{\partial s}\notag
\\&-\frac{2}{s}\frac{\partial F(p_1,p_2,p_4,p_3,s,\tilde{u})}{\partial s}.
\end{align}

\subsection{Secondary Conformal Ward Identities}
Since our 4-point function is symmetric in $\mu_1\leftrightarrow \nu_1$, the terms proportional to $\delta^{\mu_1 \kappa}$ and $\delta^{\nu_1 \kappa}$ given by the coefficients $\tilde{C}_{41}$ and $\tilde{C}_{51}$ identify a single constraint, as well as $\tilde{C}_{42}$ and $\tilde{C}_{52}$, and are explicitly given by the factors $\tilde{C}_{4j}$. They take the form
\begin{equation}
\begin{split}
&\tilde{C}_{41}=\hat{G}\Big(F(p_1,p_2,p_3,p_4,s,t)-F(p_1,p_4,p_3,p_2,t,s)\Big)+\hat{A}F(p_1,p_2,p_4,p_3,s,\tilde{u}),
\end{split}
\end{equation}
where 
\begin{equation}
\begin{split}
\hat{G}&=\frac{d(s^2+t^2-p_4^2-p_2^2-2p_1^2)+2\Delta p_1^2}{p_1^2}-\frac{t^2+p_3^2-p_2^2}{t}\frac{\partial}{\partial t}+\frac{p_2^2+p_4^2-s^2-t^2}{p_1}\frac{\partial}{\partial p_1}
\\&-\frac{s^2+p_3^2-p_4^2}{s}\frac{\partial}{\partial s}-2p_3\frac{\partial}{\partial p_3},
\end{split}
\end{equation}
and
\begin{equation}
\begin{split}
\hat{A}&=\Bigg(\frac{d(s^2+p_4^2-p_2^2-t^2)}{p_1^2}\Bigg)+\frac{t^2+p_2^2-p_3^2}{t}\frac{\partial}{\partial t}+\frac{p_2^2+t^2-p_4^2-s^2}{p_1}\frac{\partial}{\partial p_1}
\\&+\frac{p_3^2-p_4^2-s^2}{s}\frac{\partial}{\partial s}+2 p_2\frac{\partial}{\partial p_2}-2p_4\frac{\partial}{\partial p_4}.
\end{split}
\end{equation}
Moreover, we obtain
\begin{equation}
\begin{split}
&\tilde{C}_{42}=\hat{M}\Big(F(p_1,p_2,p_3,p_4,s,t)-F(p_1,p_2,p_4,p_3,s,\tilde{u})\Big)+\hat{N}F(p_1,p_4,p_3,p_2,t,s),
\end{split}
\end{equation}
where
\begin{equation}
\begin{split}
\hat{M}=&\left(\frac{2\Delta p_1^2-d(p_1^2-p_2^2+s^2)}{p_1^2}\right)-\frac{p_1^2+p_2^2-s^2}{p_1}\frac{\partial}{\partial p_1}+\frac{p_3^2-p_2^2-t^2}{t}\frac{\partial}{\partial t}-2 p_2\frac{\partial}{p_2}
\end{split}
\end{equation}
and
\begin{equation}
\begin{split}
\hat{N}=&\left(\frac{d \left(p_1^2+p_2^2+2p_4^2-s^2-2 t^2\right)}{p_1^2}\right)+\frac{t^2+p_3^2-p_2^2}{t}\frac{\partial}{\partial t}-\frac{p_1^2+p_2^2+2p_4^2-s^2-2t^2}{p_1}\frac{\partial}{\partial p_1}\\&+\frac{2 (p_3^2-p_4^2)}{s}\frac{\partial}{\partial s}-2p_4\frac{\partial}{\partial p_4}+2p_3\frac{\partial}{\partial p_3}.
\end{split}
\end{equation}

Combining  \eqref{ConsWI22} along with  \eqref{CWIlocal} and \eqref{fullform} we obtain the equations
\begin{equation}\label{example}
\begin{split}
&\hat{G}\Big(F(p_1,p_2,p_3,p_4,s,t)-F(p_1,p_4,p_3,p_2,t,s)\Big)+\hat{A}F(p_1,p_2,p_4,p_3,s,\tilde{u})=\\
&\hspace{2cm}=\frac{4d}{p_1^2}\Big(\langle O(\bold{p_1+p_2})O(\bold{p_3})O(\mathbf{\bar{p}_4})\rangle-\langle O(\bold{p_2})O(\bold{p_3})O(\bold{p_2+p_3}) \rangle \Big),
\end{split}
\end{equation}
and
\begin{equation}
\begin{split}
&\hat{M}\Big(F(p_1,p_2,p_3,p_4,s,t)-F(p_1,p_2,p_4,p_3,s,\tilde{u})\Big)+\hat{N}F(p_1,p_4,p_3,p_2,t,s)=\\
&\hspace{2cm}=\frac{4d}{p_1^2}\big(\langle O(\bold{p_2})O(\bold{p_1+p_3})O(\mathbf{\bar{p}_4})\rangle-\langle O(\bold{p_2})O(\bold{p_3})O(\bold{p_2+p_3}) \rangle\big).
\end{split}
\end{equation}
These are the secondary WI's for the $TOOO$. The 3-point function on the right hand side of this equation is uniquely given by a combination of hypergeometric functions and will be discussed below. 

\section{Asymptotics for scalar and dual conformal/conformal 4-point functions}
\label{asimpt}
Our goal, from this section on, will be to identify some of the properties of these primary and secondary equations for the $TOOO$, and for this reason it will be compelling to consider first the $(OOOO)$ correlator, which is slightly simpler compared to the former. Both cases show some similarities, starting from the fact that they are both characterised by a single form factor. The structure of the equations is expected to be similar, and indeed in both cases we will be able to identify also a similar behaviour in the corresponding form factors, in some kinematical limits. \\
The $OOOO$, as shown recently \cite{Maglio:2019grh}, allows a specific class of solutions which are uniquely identified by enlarging the original conformal symmetry to include a dual conformal symmetry as well. Indeed, these special solutions are very useful 
for studying the hypergeometric structure of the CWI's in some asymptotic limits. As we are going to see, hypergeometric solutions of 4-point functions are very special, as one expects on generic grounds, and the general CWI's, even in the scalar case, are not described by hypergeometric systems related to $F_4$. The  only exact statement that can be made concerning the structure of such systems of equations, as we are going to show, will be that Lauricella functions - i.e. hypergeometric functions of three variables - are exact solutions of all these systems of equations and can be interpreted as homogeneous (i.e. particular) solutions of such CWI's for arbitrary scaling dimensions of the scalar operators. \\  
We start our discussion by recalling that for the $OOOO$, the two CWI's take the form 
(general scalar CWI's)
\cite{Maglio:2019grh}

\begin{align}
S_1=&=\bigg\{\frac{\partial^2}{\partial p_2^2}+\frac{(d-2\D_2+1)}{p_2}\frac{\partial}{\partial p_2}-\frac{\partial^2}{\partial p_4^2}-\frac{(d-2\D_4+1)}{p_4}\frac{\partial}{\partial p_4}\notag\\
&\qquad+\frac{1}{s}\frac{\partial}{\partial s}\left(p_1\frac{\partial}{\partial p_1}+p_2\frac{\partial}{\partial p_2}-p_3\frac{\partial}{\partial p_3}-p_4\frac{\partial}{\partial p_4}\right)+\frac{\Delta_{3412}}{s}\frac{\partial}{\partial s}\notag\\
&\qquad+\frac{1}{t}\frac{\partial}{\partial t}\left(p_2\frac{\partial}{\partial p_2}+p_3\frac{\partial}{\partial p_3}-p_1\frac{\partial}{\partial p_1}-p_4\frac{\partial}{\partial p_4}\right)+\frac{\Delta_{1423}}{t}\frac{\partial}{\partial t}\notag\\[1.2ex]
&\qquad+\frac{(p_2^2-p_4^2)}{st}\frac{\partial^2}{\partial s\partial t}\bigg\}\,\Phi(p_1,p_2,p_3,p_4,s,t)=0
\label{C2}
\end{align}

\begin{align}
S_2&=\bigg\{\frac{\partial^2}{\partial p_1^2}+\frac{(d-2\D_1+1)}{p_1}\frac{\partial}{\partial p_1}-\frac{\partial^2}{\partial p_3^2}-\frac{(d-2\D_3+1)}{p_3}\frac{\partial}{\partial p_3}\notag\\
&\qquad+\frac{1}{s}\frac{\partial}{\partial s}\left(p_1\frac{\partial}{\partial p_1}+p_2\frac{\partial}{\partial p_2}-p_3\frac{\partial}{\partial p_3}-p_4\frac{\partial}{\partial p_4}\right)+\frac{\Delta_{3412}}{s}\frac{\partial}{\partial s}\notag\\
&\qquad+\frac{1}{t}\frac{\partial}{\partial t}\left(p_1\frac{\partial}{\partial p_1}+p_4\frac{\partial}{\partial p_4}-p_2\frac{\partial}{\partial p_2}-p_3\frac{\partial}{\partial p_3}\right)+\frac{\Delta_{1423}}{t}\frac{\partial}{\partial t}\notag\\[1.2ex]
&\qquad+\frac{(p_1^2-p_3^2)}{st}\frac{\partial^2}{\partial s\partial t}\bigg\}\,\Phi(p_1,p_2,p_3,p_4,s,t)=0.\label{Eq2}
\end{align}

\begin{align}
S_3&=\bigg\{\frac{\partial^2}{\partial p_1^2}+\frac{(d-2\D_1+1)}{p_1}\frac{\partial}{\partial p_1}-\frac{\partial^2}{\partial p_4^2}-\frac{(d-2\D_4+1)}{p_4}\frac{\partial}{\partial p_4}\notag\\[1.5ex]
&\qquad+\frac{1}{s}\frac{\partial}{\partial s}\left(p_1\frac{\partial}{\partial p_1}+p_2\frac{\partial}{\partial p_2}-p_3\frac{\partial}{\partial p_3}-p_4\frac{\partial}{\partial p_4}\right)+\frac{\Delta_{3412}}{s}\frac{\partial}{\partial s}\notag\\[1.5ex]
&\qquad+\frac{(p_2^2-p_3^2)}{st}\frac{\partial^2}{\partial s\partial t}\bigg\}\,\Phi(p_1,p_2,p_3,p_4,s,t)=0\label{C1}
\end{align}
where 
 \begin{equation}
 \label{deltas}
 \Delta_{ijkl}=\Delta_i +\Delta_j-\Delta_k-\Delta_l
 \end{equation}
 is a specific combination of the scaling parameters of the primary scalar operators $(O)$, which plays a special role in the derivation of the dcc solutions.
In \cite{Maglio:2019grh} the discussion dealt with two possible cases for the $OOOO$ in which the scaling combinations in \eqref{deltas} vanish: 1) the equal scaling case with $\Delta_i=\Delta$ (i=1,2,3,4) and 
2) the case in which two operators are pairwise of equal scalings. In both cases, the solutions satisfy the condition of being conformal and dual conformal invariant.\\
 The vanishing of \eqref{deltas} is necessary in order to 
remove the $\partial/\partial s$ and $\partial/\partial t$ terms and reduce the three $S_i$'s to a  hypergeometric system of equations \eqref{diff}. Notice that differently from the case of 3-point functions, where a similar system has been identified \cite{Coriano:2013jba}, as shown in Eq. \eqref{ipergio}, the variables are quartic - rather than quadratic - ratios of the invariants. \\
In order to derive such a system, which is extracted from the $S_i$'s, we need a product ans\"atz based on a quartic pivot $(s^2 t^2)$ with variables \cite{Maglio:2019grh}\cite{Bzowski:2013sza}
 \begin{equation}
 x=\frac{p_1^2 p_3^2}{s^2 t^2}\qquad  y=\frac{p_2^2 p_4^2}{s^2 t^2},
 \label{xy}
 \end{equation}
and  observe that this choice sets automatically to zero the mixed derivative terms in $p_i$ and $s$ and $t$ in Eqs. \eqref{C2},\eqref{Eq2} and \eqref{C1}. The ans\"atz for the solution is based on the product of a function G(x,y) and of powers of x and y - given by \eqref{xy} - of the form 
\begin{equation} 
\Phi\sim x^a y^b G(x,y),
\end{equation}
for suitable $a$ and $b$, quite similarly to the case of a scalar 3-point function.
On any function $G(x,y)$, terms of the form

\begin{equation}
\left( p_1\frac{\partial}{\partial p_1}+p_2\frac{\partial}{\partial p_2}-p_3\frac{\partial}{\partial p_3}-p_4\frac{\partial}{\partial p_4}\right)G(x,y)=0
\label{GG}
\end{equation}
vanish, if we choose $x$ and $y$ as the quartic ratios \eqref{xy}.
If we use the definition of the $K_{ij}$ operators  \eqref{kappa}, \eqref{ipergio} and the ans\"atz based on $G(x,y)$ as defined above, the three equations take the form (intermediate scalar CWI's)
\begin{equation}
\left\{
\begin{split}
&\bigg( K_{2 4} +\frac{(p_2^2-p_4^2)}{st}\frac{\partial^2}{\partial s\partial t}\bigg)\Phi(p_1,p_2,p_3,p_4,s,t)=-\bigg(\frac{\Delta_{3412}}{s}\frac{\partial}{\partial s}
+\frac{\Delta_{1423}}{t}\frac{\partial}{\partial t} \bigg)\,\Phi(p_1,p_2,p_3,p_4,s,t)
 \\ 
&\bigg( K_{1 3} +\frac{(p_1^2-p_3^2)}{st}\frac{\partial^2}{\partial s\partial t}\bigg)\Phi(p_1,p_2,p_3,p_4,s,t)=-\bigg(\frac{\Delta_{3412}}{s}\frac{\partial}{\partial s}
+\frac{\Delta_{1423}}{t}\frac{\partial}{\partial t} \bigg)\,\Phi(p_1,p_2,p_3,p_4,s,t)\\ 
&\bigg( K_{1 4} +\frac{(p_2^2-p_3^2)}{st}\frac{\partial^2}{\partial s\partial t}\bigg)\Phi(p_1,p_2,p_3,p_4,s,t)=-\frac{\Delta_{3412}}{s}\frac{\partial}{\partial s} \,\Phi(p_1,p_2,p_3,p_4,s,t)\\
\label{CC4}
\end{split}
\right.\\
\end{equation}
where we have removed all the mixed derivative terms in $(s, p_i^2),(t, p_i^2)$, thanks to  \eqref{GG}. Explicit dcc solutions of this system of equations are obtained if $\Delta_{ijkl}=0$, and the operators 
$K_{24}$ and $K_{13} $ depend separately on a {\em single scaling variable}, that is if 
$\Delta_2=\Delta_4$ and $\Delta_1=\Delta_3$. Notice that this condition is compatible with the vanishing of $\Delta_{3412}$ and $\Delta_{1423}$ and takes to a hypergeometric system of equations, which are again solved in terms of hypergeometrics of the variables $x$ and $y$ given in \eqref{xy}. In this case we could rewrite the system in the form (reduced scalar CWI's)
\begin{equation}
\left\{
\begin{split}
&\bigg( K_{2 4}(\Delta_2) +\frac{(p_2^2-p_4^2)}{st}\frac{\partial^2}{\partial s\partial t}\bigg)\Phi(p_1,p_2,p_3,p_4,s,t)=0
\\ 
&\bigg( K_{1 3}(\Delta_1) +\frac{(p_1^2-p_3^2)}{st}\frac{\partial^2}{\partial s\partial t}\bigg)\Phi(p_1,p_2,p_3,p_4,s,t)=0 \\
&\bigg( K_{1 4} +\frac{(p_2^2-p_3^2)}{st}\frac{\partial^2}{\partial s\partial t}\bigg)\Phi(p_1,p_2,p_3,p_4,s,t)=0
\label{CCC4}
\end{split}
\right.\\
\end{equation}
where the $K_{ij}(\Delta_i)$ indicates that such operators depend on a single scaling constant. \\
It is important to observe that the system \eqref{CCC4} admits explicit dcc solutions which are expressed as 
hypergeometric functions, or, equivalently, as 3K integrals, but the entire set of dcc solutions is not just composed of these functions. We refer to appendix E for few comments on the properties of such solutions.\\ 
Dual conformal symmetry constrains a certain ans\"atz (the dual conformal ans\"atz) to be expressed only in terms of the two quartic ratios $x$ and $y$, via a function $G(x,y)$. Functions $G$ of such ratios will then necessarily satisfy the condition \eqref{GG}, and henceforth the reduced system \eqref{CC4}. \\
The solutions of the three constraints in \eqref{CC4} of the form $G(x,y)$, will then characterize the most general set of dcc solutions for scalar primary operators, of which special cases are those found in \cite{Maglio:2019grh} and reported below in Eq. \eqref{fform}.
The additional reduction of the system \eqref{CC4} to \eqref{CCC4} obviously, allows us to work with explicit expressions which are all related by analytic continuations and therefore describe a unique solution, as shown in \cite{Maglio:2019grh}. Therefore, they are optimal for the study of several kinematical limits of the scalar correlator, that we are now going to investigate.

\subsection{Limits for equal scalings and $\Delta_{ijkl}=0$}
As we have mentioned, the choice $\Delta_{ijkl}=0$ is what renders the system \eqref{CCC4} a variant of the ordinary hypergeometric system, which in general takes the form \eqref{ipergio} and it is solved by quadratic - rather than quartic - ratios of invariants.  
Once this gets reduced to \eqref{CCC4},  as already mentioned, the complete ans\"atz for the general solution of such system is constructed by multiplying the function $G(x,y)$ by the pivot, raised to a power $n_s$, fixed by the dilatation WI  
\begin{align}
\bigg[(\D_t-3d)-\sum_{i=1}^4p_i\frac{\partial}{\partial p_i}-s\frac{\partial}{\partial s}-t\frac{\partial}{\partial t}\bigg]\Phi(p_1,p_2,p_3,p_4,s,t)=0,\label{Dilatation4}
\end{align}
with $\Delta_t$ denoting the total scaling. If we choose as a pivot $s^2 t^2$,
 the solution indeed will take the form
\begin{equation}
 \Phi(p_i,s,t)= (s^2 t^2)^{n_s}G(x,y) \qquad  n_s=\frac{\Delta_t -3 d}{4}.
 \label{ans}
\end{equation}
Few additional comments are in order concerning the homogeneous case 
$(\Delta_{ijkl}=0)$ and the system \eqref{CCC4}. We remark that the third equation of such system is identically satisfied if the first and the second equations are, which is the case if an ans\"atz of type \eqref{ans} is chosen. This is clearly consistent with the fact the four functionally independent solutions of an Appell system of equations (for $F_4$) is based only on two independent equations \eqref{diff}.
 
 The solution of the homogeneous system \eqref{CCC4}, as already mentioned, can be written in terms of 4 Appell functions $F_4$ of the $x$ and $y$ ratios given in \eqref{xy} 
 \cite{Maglio:2019grh}
\begin{align}
\braket{O(p_1)O(p_2)O(p_3)O(p_4)}&=2^{\frac{d}{2}-4}\ \ C\,\sum_{\l,\m=0,\D-\frac{d}{2}}\x(\l,\m)\bigg[\big(s^2\,t^2\big)^{\D-\frac{3}{4}d}\left(\frac{p_1^2 p_3^2}{s^2 t^2}\right)^\l\left(\frac{p_2^2p_4^2}{s^2t^2}\right)^\m\nonumber\\
&\hspace{-3cm}\times\,F_4\left(\frac{3}{4}d-\D+\l+\m,\frac{3}{4}d-\D+\l+\m,1-\D+\frac{d}{2}+\l,1-\D+\frac{d}{2}+\m,\frac{p_1^2 p_3^2}{s^2 t^2},\frac{p_2^2 p_4^2}{s^2 t^2}\right)\notag\\
&+\big(s^2\,u^2\big)^{\D-\frac{3}{4}d}\left(\frac{p_2^2 p_3^2}{s^2 u^2}\right)^\l\left(\frac{p_1^2p_4^2}{s^2u^2}\right)^\m\notag\\
&\hspace{-3cm}\times\,F_4\left(\frac{3}{4}d-\D+\l+\m,\frac{3}{4}d-\D+\l+\m,1-\D+\frac{d}{2}+\l,1-\D+\frac{d}{2}+\m,\frac{p_2^2 p_3^2}{s^2 u^2},\frac{p_1^2 p_4^2}{s^2 u^2}\right)\notag\\
&+\big(t^2\,u^2\big)^{\D-\frac{3}{4}d}\left(\frac{p_1^2 p_2^2}{t^2 u^2}\right)^\l\left(\frac{p_3^2p_4^2}{t^2u^2}\right)^\m\notag\\
&\hspace{-3cm}\times\,F_4\left(\frac{3}{4}d-\D+\l+\m,\frac{3}{4}d-\D+\l+\m,1-\D+\frac{d}{2}+\l,1-\D+\frac{d}{2}+\m,\frac{p_1^2 p_2^2}{t^2 u^2},\frac{p_3^2 p_4^2}{t^2 u^2}\right)\bigg],\label{fform}
\end{align}
where the coefficients $\x(\l,\m)$ are explicitly given by
\begin{equation}
\begin{split}
\x\left(0,0\right)&=\left[\Gamma\left(\frac{3}{4}d-\D\right)\right]^2\left[\Gamma\left(\D-\frac{d}{2}\right)\right]^2\\
\x\left(0,\D-\frac{d}{2}\right)&=\x\left(\D-\frac{d}{2},0\right)=\left[\Gamma\left(\frac{d}{4}\right)\right]^2\Gamma\left(\D-\frac{d}{2}\right)\Gamma\left(\frac{d}{2}-\D\right)\\
\x\left(\D-\frac{d}{2},\D-\frac{d}{2}\right)&=\left[\Gamma\left(\D-\frac{d}{4}\right)\right]^2\left[\Gamma\left(\frac{d}{2}-\D\right)\right]^2,
\end{split}\label{xicoef}
\end{equation}
which is explicitly symmetric under all the possible permutations of the momenta and it is fixed up to one undetermined constant $C$.\\
As shown in \cite{Maglio:2019grh}, \eqref{fform} can be re-expressed in the form  
\begin{align}
&I_{\frac{d}{2}-1\{\Delta-\frac{d}{2},\Delta-\frac{d}{2},0\}}(p_1p_3,p_2 p_4,s t)=\notag\\
&\qquad=\,(p_1p_3)^{\Delta-\frac{d}{2}}(p_2p_4)^{\Delta-\frac{d}{2}}\int_{0}^\infty\,dx\,x^{\frac{d}{2}-1}\,K_{\Delta-\frac{d}{2}}(p_1p_3\,x)\,K_{\Delta-\frac{d}{2}}(p_2 p_4\,x)\,K_{0}(st\,x).\label{Sol}
\end{align}
i.e. as a 3K integrals of quadratic $(p_1 p_3, s t, p_2 p_4)$ variables, which are solutions of a system of the form \eqref{diff} with quartic ratios $x, y$.
Few technical details are given in appendix \eqref{diff}.
\subsection{Comparison between the general, the intermediate and the reduced systems}

To address the asymptotic behaviour of this solution of the general system of Eqs. 
\eqref{C2}\eqref{Eq2}\eqref{C1} (the $S_i$ constraints) and compare it with the intermediate 
\eqref{CC4} and the reduced \eqref{CCC4} ones, we clearly need to perform a special asymptotic limit. We can reasonably assume that at large $s$ and $t$ the general solution of the $S_{i}'s$ equations decays as $\sim 1/(s t)^\alpha$, with $\alpha > 0$. \\
 Both for the $S_i$ and for the intermediate system \eqref{CC4}, the action of the derivative operators $(1/s) \partial/\partial s$ and $(1/t) \partial/\partial t$ is suppressed by two additional powers of the kinematic invariant $s$ and $t$ and can reasonably be set to zero asymptotically.\\
 If we neglect such contributions, the equations in \eqref{CC4} turn again into a homogeneous system  \eqref{fform} which, however, is not hypergeometric any longer, nor the third equation is dependent from the previous two, as found in the $\Delta_{ijkl}=0$ case for the reduced system \eqref{CCC4}.  Although the three systems, general intermediate and reduced, look pretty similar in such limit, we can only safely state that their solutions have to share the same asymptotic behaviour. This is fixed by the scaling power 
 $n_s=\Delta_t -3/4 d $, extracted from the dilatation WI in the form  
\begin{equation}
\Phi(p_1,p_2,p_3,p_4)\sim\frac{1}{(s^2 t^2)^{- n_s}} + O(1/(s^2 t^2))
\end{equation}
which requires that $n_s$ be negative. \\
In the two sections below we will try to characterize the behaviour of the dcc solution of \eqref{CCC4} in various limits before coming back again to the three systems of equations, discussing some approximate factorised solutions of such equations.

\subsection{IR and equal mass limits of the dcc solutions}
The analysis of the infrared or soft limits at small $s$ and $t$ of the dual conformal solution, with $\Delta_{ijkl}=0$, for $\Delta_i=\Delta, i=1,2,3,4$ can be discussed using a second version of the solution given by 
\eqref{fform}, but completely equivalent to it, obtained by a sequence of analytic continuations  \cite{Maglio:2019grh}

\begin{equation}
\resizebox{0.6\hsize}{!}{$
\begin{aligned}
\Phi&=C_1\bigg\{\left(p_1^2\,p_3^2\right)^{\D-\frac{3}{4}d}\bigg[F_4\left(\frac{d}{4}\,,\,\frac{3}{4}d-\D\,,\,1\,,\,\frac{d}{2}-\D+1\,;\frac{s^2t^2}{p_1^2p_3^2}\,,\,\frac{p_2^2p_4^2}{p_1^2p_3^2}\right)\\[1.2ex]
&\hspace{-0.3cm}+\tau_1\left(\frac{p_2^2p_4^2}{p_1^2p_3^2}\right)^{\D-\frac{d}{2}} F_4\left(\D-\frac{d}{4}\,,\,\frac{d}{4}\, ,\,1\,,\,1-\frac{d}{2}+\D\,;\frac{s^2t^2}{p_1^2p_3^2}\,,\,\frac{p_2^2p_4^2}{p_1^2p_3^2}\right)\bigg]\\[1.2ex]
&+\left(p_2^2\,p_3^2\right)^{\D-\frac{3}{4}d}\bigg[F_4\left(\frac{d}{4}\,,\,\frac{3}{4}d-\D\, ,\,1\,,\,\frac{d}{2}-\D+1\,;\frac{s^2u^2}{p_2^2p_3^2}\,,\,\frac{p_1^2p_4^2}{p_2^2p_3^2}\right)\\[1.2ex]
&\hspace{-0.3cm}+\tau_1 \left(\frac{p_1^2p_4^2}{p_2^2p_3^2}\right)^{\D-\frac{d}{2}} F_4\left(\D-\frac{d}{4}\,,\,\frac{d}{4}\, ,\,1\,,\,1-\frac{d}{2}+\D\,;\frac{s^2u^2}{p_2^2p_3^2}\,,\,\frac{p_1^2p_4^2}{p_2^2p_3^2}\right)\bigg]\\[1.2ex]
&+\left(p_1^2\,p_2^2\right)^{\D-\frac{3}{4}d}\bigg[F_4\left(\frac{d}{4}\,,\,\frac{3}{4}d-\D\, ,\,1\,,\,\frac{d}{2}-\D+1\,;\frac{u^2t^2}{p_1^2p_2^2}\,,\,\frac{p_3^2p_4^2}{p_1^2p_2^2}\right)\\[1.2ex]
&\hspace{-0.3cm}+\tau_1 \left(\frac{p_3^2 p_4^2}{p_1^2 p_2^2}\right)^{\Delta -\frac{d}{2}}F_4\left(\D-\frac{d}{4}\,,\,\frac{d}{4}\, ,\,1\,,\,1-\frac{d}{2}+\D\,;\frac{u^2t^2}{p_1^2p_2^2}\,,\,\frac{p_3^2p_4^2}{p_1^2p_2^2}\right)\bigg]\bigg\}.
\end{aligned}$}
\label{twoo}
\end{equation}

\begin{equation}
\tau_1=\frac{\Gamma\left(\D-\frac{d}{4}\right)\Gamma\left(1+\D-\frac{3}{4}d\right)\Gamma\left(1-\D+\frac{d}{2}\right)}{\Gamma\left(\D-\frac{3}{4}d\right)\Gamma\left(1-\D+\frac{3}{4}d\right)\Gamma\left(1+\D-\frac{d}{2}\right)}
\end{equation}
From this expression, we can keep $t^2$ fixed and of order $O(p_i^2)$ and send $s^2\to 0 $ to derive the soft behaviour 
\begin{equation}
\Phi\sim (p_1^2 p_3^2)^{\Delta - \frac{3}{4}d} + O(s^2/p_i^2)
\end{equation}
if the external mass invariants $p_i^2$ are kept larger than the invariant $s^2$. 
\subsection{Equal mass limit with $p_i^2=M^2 >\, s^2,t^2$ }
The equal mass limit is obtained by taking $p_i^2=M^2$ for all the external invariants. In this case, using the relation between $F_4$ and the Gauss hypergeometric $F_{2 1}(a,b,c,x)$
\begin{equation}
F_4(\alpha,\beta,\gamma,\gamma',x,y)=\sum_{m=0}^{\infty}\frac{(\alpha)_m(\beta)_m}
{(\gamma)_m m!}F_{21}(\alpha + m,\beta +m,\gamma',y)x^m
\end{equation}
and 

\begin{equation} 
F_{21}(a,b,c,1)=\frac{\Gamma(c)\Gamma(c-a-b)}{\Gamma(c-a)\Gamma(c-b)},
\end{equation}
from \eqref{twoo} we then obtain the simplified expression 
\begin{equation}
\begin{split}
&\Phi=M^{4 \Delta-3d }\sum_{m=0}^\infty \frac{1}{m!} \Bigg(
\frac{\Gamma \left(\frac{d}{2}-\Delta +1\right) \Gamma
   \left(-\frac{d}{2}-2 m+1\right) \Gamma
   \left(\frac{d}{4}+m\right) \Gamma \left(\frac{3
   d}{4}-\Delta +m\right)}{ \Gamma
   \left(\frac{d}{4}\right) \Gamma (m+1) \Gamma
   \left(\frac{3 d}{4}-\Delta \right) \Gamma
   \left(-\frac{d}{4}-m+1\right) \Gamma
   \left(\frac{d}{4}-\Delta -m+1\right)} \nonumber \\
&\frac{\Gamma \left(\frac{d}{2}-\Delta +1\right) \Gamma
   \left(-\frac{3 d}{4}+\Delta +1\right) \Gamma
   \left(-\frac{d}{2}-2 m+1\right) \Gamma
   \left(\frac{d}{4}+m\right) \Gamma
   \left(-\frac{d}{4}+\Delta +m\right)}{\Gamma
   \left(\frac{d}{4}\right) \Gamma (m+1) \Gamma
   \left(\frac{3 d}{4}-\Delta +1\right) \Gamma
   \left(\delta -\frac{3 d}{4}\right) \Gamma
   \left(-\frac{d}{4}-m+1\right) \Gamma \left(-\frac{3
   d}{4}+\Delta -m+1\right)} \Bigg)\times \nn\\
 &  \times \Bigg(\left(\frac{s^2 t^2}{M^4}\right)^m + \left(\frac{s^2 u^2}{M^4}\right)^m + 
 \left(\frac{u^2 t^2}{M^4}\right)^m\Bigg)\\
   \end{split}
\end{equation}
which in $d=4$ becomes
\begin{equation}
\begin{split}
&\Phi=M^{4 \Delta-3d}\sum_{m=0}^\infty \frac{1 }{m!}  \left(\frac{\Gamma (-2 m-1) \Gamma (-\Delta
   +m+3)}{\Gamma (-m) \Gamma (-\Delta
   -m+2)}+\frac{\Gamma (3-\Delta ) \Gamma (\Delta -2)
   \Gamma (-2 m-1) \Gamma (\Delta +m-1)}{ \Gamma
   (4-\Delta ) \Gamma (\Delta -3) \Gamma (-m) \Gamma
   (\Delta -m-2)}\right)\times \\ \nonumber 
&\qquad \qquad    \times \Bigg(\left(\frac{s^2 t^2}{M^4}\right)^m + \left(\frac{s^2 u^2}{M^4}\right)^m + \left(\frac{u^2 t^2}{M^4}\right)^m\Bigg)\\
\end{split}
\end{equation}
and is convergent as far as $M^2 \gg s^2, t^2 $.  Explicitly (in $d=4$)

\begin{equation} 
\Phi =M^{4 \Delta-3d}\left[d_0 + d_1 (\frac{s^2 t^2}{M^4} + \frac{s^2 u^2}{M^4} + \frac{u^2 t^2}{M^4})+\ldots \right],
\end{equation}
where
\begin{equation}
d_0=\frac{1}{2} \left(\frac{\Gamma (\Delta -1)}{\Gamma
   (\Delta -2)}-\frac{\Gamma (3-\Delta )}{\Gamma
   (2-\Delta )}\right)
\end{equation}
and 
\begin{equation}
\begin{split}
& d_1=\frac{1}{12} \left(\frac{\Gamma (4-\Delta )}{\Gamma
   (1-\Delta )}-\frac{\Gamma (\Delta )}{\Gamma (\Delta
   -3)}\right).
\end{split}
\end{equation}
\subsection{The equal mass limit with $s^2, t^2 > M^2$} 
A similar limit can be performed starting from \eqref{fform}. We can take $s^2, t^2, u^2 > M^2$, which in \eqref{fform} takes to a univariate expression of $F_4$, $F_4(a,b,c,c';x,x)$. It can be expressed as a single series in $x$ using the relation 

\begin{equation}
F_4(a,b,c,c';x,x)= {}_4 F_{3}\left(a,b,\frac{c + c'}{2},\frac{c + c'-1}{2}| \,c,c', c + c'-1; 4 x\right)  
\end{equation}
due to Burchnall, as reported in \cite{Vidunas1}. \\
Setting $x_1=M^4/(s^2 t^2)\sim M^4/(s^2 u^2)\sim M^4/(u^2 t^2)$ and choosing, for instance, a scaling dimension of a scalar operator 
$\phi^2$ (with $\Delta=d-2$), in $d=4$ one obtains a simple expression for $\Phi$

\begin{equation}
\Phi=C\left(\frac{\log ^2\left(\frac{M^4}{s^2 t^2}\right)}{s^2
   t^2}+\frac{\log ^2\left(\frac{M^4}{s^2
   u^2}\right)}{s^2 u^2}+\frac{\log
   ^2\left(\frac{M^4}{t^2 u^2}\right)}{t^2
   u^2}+\frac{\pi ^2}{3 s^2 t^2}+\frac{\pi ^2}{3 s^2
   u^2}+\frac{\pi ^2}{3 t^2 u^2}\right) +O(x_1^2)
\end{equation}
and in $d=3$
\begin{equation}
\begin{split}
&\Phi=C\left(\frac{\pi  \Gamma \left(\frac{1}{4}\right)^2}{M^4
   \sqrt{s} \sqrt{t}}+\frac{\pi  \Gamma
   \left(\frac{1}{4}\right)^2}{M^4 \sqrt{s}
   \sqrt{u}}+\frac{\pi  \Gamma
   \left(\frac{1}{4}\right)^2}{M^4 \sqrt{t}
   \sqrt{u}}-\frac{4 \pi  \Gamma
   \left(\frac{3}{4}\right)^2}{M^2 s^{3/2}
   t^{3/2}}-\frac{4 \pi  \Gamma
   \left(\frac{3}{4}\right)^2}{M^2 s^{3/2}
   u^{3/2}}\right.\nn \\
 &\qquad \qquad  \left. -\frac{4 \pi  \Gamma
   \left(\frac{3}{4}\right)^2}{M^2 t^{3/2}
   u^{3/2}}+\frac{\pi  \Gamma
   \left(\frac{1}{4}\right)^2}{4 s^{5/2}
   t^{5/2}}+\frac{\pi  \Gamma
   \left(\frac{1}{4}\right)^2}{4 s^{5/2}
   u^{5/2}}+\frac{\pi  \Gamma
   \left(\frac{1}{4}\right)^2}{4 t^{5/2} u^{5/2}}\right) +O(x_1^2)
\end{split}
\end{equation} 

\section{Large $s$ and $t$ limits and the Lauricella system}
We have already mentioned that the system of Eqs. \eqref{C2}, \eqref{Eq2} \eqref{C1} reduces to \eqref{CC4} if we choose a combination of invariants given by \eqref{xy}. 
The system \eqref{CC4} turns into hypergeometric if $\Delta_{ijkl}=0$, with only two independent equations, as pointed out above. However, for a generic $\Delta_{ijkl}$ it is possible to uncover an approximate hypergeometric structure in the equations only in the large $s$ and $t$ limit, if we neglect the coupling between the $s, t$ and $p_i^2$ invariants. At the same time
we could assume that $\Delta_{ijkl}\ll1$, which allows to drop the $1/s\partial/\partial_s$ and  $1/t\partial/\partial_t$ terms in the differential operator. This approximate factorization has been discussed in \cite{Maglio:2019grh}, where it has been shown to take to a hypergeometric system of Lauricella type in three variables (see section \eqref{lauri}). This asymptotic analysis is based on the ans\"atz

\begin{equation}
\Phi(p_1^2,p_2^2,p_3^2,p_4^2,s, t)\sim \phi(p_1^2,p_2^2,p_3^2,p_4^2)\chi(s,t)
\label{factor}
\end{equation}
and invokes the separability of the asymptotic system \eqref{CC4} 
\begin{equation}
\label{lau}
\begin{split}
&K_{2 4} \phi=0 \\
& K_{1 3}\phi=0 \\
& K_{1 4}\phi=0\\ 
&\frac{1}{st}\frac{\partial^2}{\partial s\partial t}\,\chi(s,t)=0.\\
\end{split}
\end{equation}
The Lauricella system corresponds to the first three equations of \eqref{lau}. 
Lauricella systems have recently appeared also in CFT in coordinate space \cite{Chen:2019gka}.
One can easily realize that they characterize a homogenous solution in the variables $p_i^2$ of the entire (complete) system \eqref{C2}, \eqref{Eq2}\eqref{C1} as well as of \eqref{CC4}. They are exact solutions of such systems, before an asymptotic limit. Notice that logarithmic terms of the form  $f(1/(s^2 t^2)) \log^k(s^2/t^2)$ (k>0) are also compatible with the asymptotic structure of such systems, which are generically expected for scattering at fixed angle in perturbation theory (see \cite{Cheng:1982xq} for an example). 
 
\subsection{The general primary CWI's in the $2\to 2$  and $1\to 3$ formulations for the $TOOO$ and asymptotics} 
\label{gen}
In order to get further insight into the structure of the CWI's for the $TOOO$, we proceed with a rearrangement of their expressions in order to reduce them to homogeneous equations, following the same approach of section \ref{asimpt}, adopted in the scalar case. We will proceed by generalizing the CWI's for different scalar coefficients from the equal scaling case presented in section \ref{primd}.
For simplicity we set 
\begin{equation}
F\equiv F(p_1,p_2,p_3,p_4,s, t), \,\,\, F(p_2\leftrightarrow p_4)\equiv F(p_1,p_2,p_4,p_3, t ,s),
\,\,\, F(p_3\leftrightarrow p_4)\equiv F(p_1,p_4,p_2,p_3, s,\tilde{u}).
\end{equation}
If we allow for different scaling $\Delta_i$, with $\Delta_1=d$ for the stress-energy tensor, then the equations given in \eqref{c11}-\eqref{c31} can be generalized as follows
\begin{equation}
\begin{split}
 \tilde{C}_{1 1} -\tilde{C}_{2 1}\to B_1=& \Bigg( K_{21} +\frac{\Delta_{1423}+2}{t}\frac{\partial}{\partial t} -\frac{1}{t}\frac{\partial}{\partial t}\left( p_1\frac{\partial}{\partial p_1}+
p_4\frac{\partial}{\partial p_4} - p_2\frac{\partial}{\partial p_2} - p_3\frac{\partial}{\partial p_3}\right) \\&+ \frac{p_3^2-p_4^2}{s t}\frac{\partial^2}{\partial s \partial t}\Bigg)F(1,2,3,4)=0.
\end{split}
\end{equation}

The other homogeneous equations for $F(p_1,p_2,p_3,p_4,s, t)$ are similarly derived in the form
\begin{align}
 \tilde{C}_{1 1} -\tilde{C}_{3 1}\to B_2=& \Bigg( K_{31} +\frac{\Delta_{1423}+2}{t}\frac{\partial}{\partial t}+\frac{\Delta_{1234}-2}{s}\frac{\partial}{\partial s} -\frac{1}{t}\frac{\partial}{\partial t}\left( p_1\frac{\partial}{\partial p_1}+
p_4\frac{\partial}{\partial p_4} - p_2\frac{\partial}{\partial p_2} - p_3\frac{\partial}{\partial p_3}\right)\nonumber\\
&+\frac{1}{s}\frac{\partial}{\partial s}\left( p_3\frac{\partial}{\partial p_3}+
p_4\frac{\partial}{\partial p_4} - p_1\frac{\partial}{\partial p_1} - p_2\frac{\partial}{\partial p_2}\right) + \frac{p_3^2-p_1^2}{s t}\frac{\partial^2}{\partial s \partial t}\Bigg)F(1,2,3,4)=0, 
\end{align}
and
\begin{align}
\tilde{C}_{2 1} -\tilde{C}_{3 1}\to B_3=& \Bigg( K_{32} +\frac{\Delta_{1234}-2}{s}\frac{\partial}{\partial s} +\frac{1}{s}\frac{\partial}{\partial s}\left( p_3\frac{\partial}{\partial p_3}+
p_4\frac{\partial}{\partial p_4} - p_1\frac{\partial}{\partial p_1} - p_2\frac{\partial}{\partial p_2}\right) \\& +\frac{p_4^2-p_1^2}{s t}\frac{\partial^2}{\partial s \partial t}\Bigg)F(1,2,3,4)=0.
\end{align}
One can show that $B_1,B_2,B_3$ are not independent, in fact
\begin{equation}
B_1+B_3=B_2,
\end{equation}
indicating that there are only two independent homogeneous equations involving the $F$ form factor. \\
Finally, one has to consider the system of three differential equations, composed of ($B_1,B_2)$ together with the analogous of $\tilde{C}_{21}$, given in \eqref{c21}, now for different $\Delta_i$'s, which can be written as
\begin{equation}
\left\{
\begin{split}
&\Bigg( K_{31} -\frac{1}{t}\frac{\partial}{\partial t}\left( p_1\frac{\partial}{\partial p_1}+
p_4\frac{\partial}{\partial p_4} - p_2\frac{\partial}{\partial p_2} - p_3\frac{\partial}{\partial p_3}\right)+\frac{1}{s}\frac{\partial}{\partial s}\left( p_3\frac{\partial}{\partial p_3}+
p_4\frac{\partial}{\partial p_4} - p_1\frac{\partial}{\partial p_1} - p_2\frac{\partial}{\partial p_2}\right)  \notag \\& + \frac{p_3^2-p_1^2}{s t}\frac{\partial^2}{\partial s \partial t}\Bigg)F=-\left(\frac{\Delta_{1423}+2}{t}\frac{\partial}{\partial t}+\frac{\Delta_{1234}-2}{s}\frac{\partial}{\partial s}\right)F\\[2ex]
&\Bigg[K_{42}+\frac{1}{s}\frac{\partial}{\partial s}\left( p_3\frac{\partial}{\partial p_3}+
p_4\frac{\partial}{\partial p_4} - p_1\frac{\partial}{\partial p_1} - p_2\frac{\partial}{\partial p_2}\right)+\frac{1}{t}\frac{\partial}{\partial t}\left( p_1\frac{\partial}{\partial p_1}+
p_4\frac{\partial}{\partial p_4} - p_2\frac{\partial}{\partial p_2} - p_3\frac{\partial}{\partial p_3}\right)\notag \\&+\frac{p_4^2-p_2^2}{st}\frac{\partial^2}{\partial s \partial t}\Bigg]F 
=\left(\frac{\Delta_{1423}+2}{t}\frac{\partial}{\partial t}-\frac{\Delta_{1234}}{s}\frac{\partial}{\partial s}\right)F-\frac{2}{s}\frac{\partial }{\partial s}\big(F(p_2\leftrightarrow p_4)-F(p_3\leftrightarrow p_4)\big)\\[2ex]
&\left( K_{21}  -\frac{1}{t}\frac{\partial}{\partial t}\left( p_1\frac{\partial}{\partial p_1}+
p_4\frac{\partial}{\partial p_4} - p_2\frac{\partial}{\partial p_2} - p_3\frac{\partial}{\partial p_3}\right) + \frac{p_3^2-p_4^2}{s t}\frac{\partial^2}{\partial s \partial t}\right)F=-\frac{\Delta_{1423}+2}{t}\frac{\partial}{\partial t}F
\end{split}
\right.\label{SystemOfHypergeometic}
\end{equation}
We will try to extract some information about the structure of such equations by discussing some possible limits.\\
In the characterization of the nature of the system we begin by considering the case in which all the scalings are different and work our way starting from the left hand side of 
\eqref{SystemOfHypergeometic}. We have different options. For instance, if we are looking for factorised solutions such as those discussed in the scalar case, of the form \eqref{factor}, 
then we could consider the asymptotic limit  $s,t\to\infty$ and identify the Lauricella component  of such solutions, since the equations above turn homogenous, and the left hand side, exactly as in \eqref{lau}, reduces to a Lauricella system of hypergeometrics \eqref{lauri}. This holds independently of the values of the scalings $\Delta_i$.
On the other hand, it is possible to identify, at least asymptotically, some hypergeometric solutions, different from the Lauricella's, but we need to constrain the scaling dimensions in such a way that the operators $K_{31}$ and $K_{42}$ are each characterised by a single conformal scaling  ($\Delta_4=\Delta_2$ and $\Delta_3=\Delta_1$). As discussed in the previous sections, we can choose as variable the scale invariant ratios \eqref{xy} for $x$ and $y$ in the ans\"atz for the solution, reobtaining the same left hand side of \eqref{CC4}. This approximate solutions would again be quite similar to those discussed in the scalar case.  
However, inn the general case, as we have already mentioned, even for large $s$ and $t$, when we keep the scalings generic, one can show that the left hand side of such system of equations is not of hypergeometric form, and the explicit form of such solutions is unknown.

\begin{figure}[t]
	\centering
	\raisebox{-0.5\height}{\includegraphics[scale=0.8]{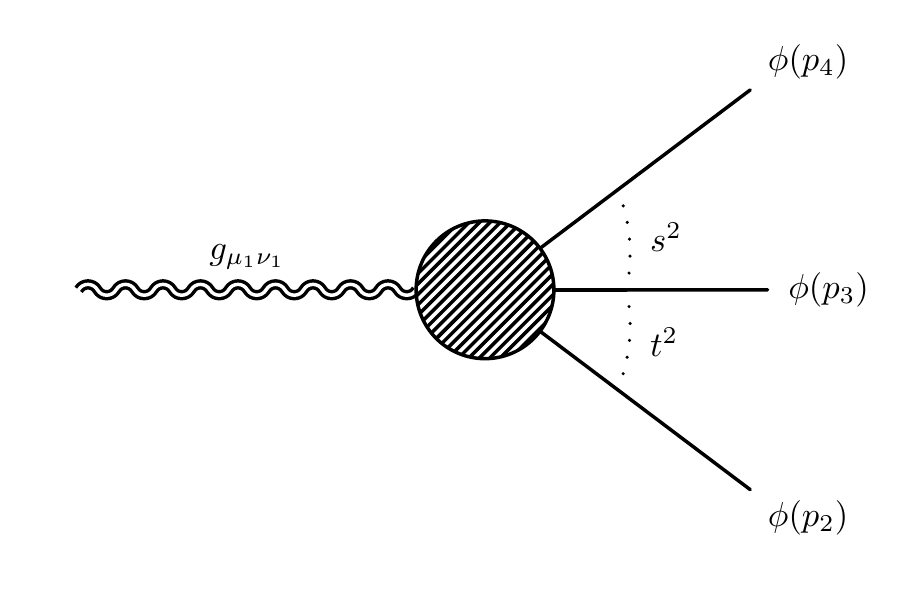}} \hspace{2cm}
	
	\caption{The $TOOO$ in a kinematical region in which can be described as a 
		$1\to 3$ process. }
	\label{nm1}
\end{figure} 
\subsection{The $1\to 3$ case} 
It is possible to perform other limits on the same form factor of the $TOOO$ in order to simplify the primary CWI's presented in the previous sections. We are going to focus our discussion on the $1\to 3$ formulation, which is symmetric in the momenta of the three scalar operators and provides a clear separation of the parametric dependence of the correlator in terms of a function of the external invariants $p_i^2$ times a function of $s,t$ and $u$, in analogy with the discussion presented in the $2\to 2$ case. \\
This kinematic choice is illustrated in Fig. 1.\\
In order to proceed with the investigation of this limit, it is convenient to perform an analytic continuation of the CWI's to the Minkowski region from their Euclidean definition, and take  all the invariants $t^2$ and $u^2$ and $s^2$ to be positive.  The kinematical region of interest, in this case, is delimited by the conditions
\bea
&& (p_2 + p_3)^2\leq t^2 \leq \left( p_1 - p_4\right)^2\nonumber \\
&& (p_3 + p_4)^2\leq s^2 \leq \left( p_1 - p_2\right)^2 \nonumber\\
&& (p_2+p_4)^2 \leq u^2 \leq \left(p_1-p_3\right)^2, 
\eea
with the usual relation
\bea�
&& s^2 + t^2 + u^2= p_1^2 + p_2^2 + p_3^2 + p_4^2. 
\eea
We will be performing the large $p_1$ limit, where the invariant mass of the 
virtual graviton line gets asymptotically large, and assume that the invariants $s^2\sim t^2\sim u^2\sim p_1^2$ grow large with $p_1^2$. In this limit the primary CWI's simplify, and the equations become approximately separable in their dependence on the external $p_i^2 (i=2,3,4)$ and the remaining $(s, t, u)$ invariants. For this reason we choose asymptotic solutions of the form 
\begin{equation}
A(p_2,p_3,p_4,s,t,u)\sim \Phi(p_2,p_3,p_4)\chi(s,t,u).
\end{equation}
We study now the form of the $\chi(s,t,u)$. The corresponding equations for the $(s,t,u)$ invariants, from the primary conformal WI's, take the form
\begin{equation}
\frac{\partial^2 \chi}{\partial s\partial t } =0, \qquad  \frac{\partial^2 \chi}{\partial s \partial u } =0, \qquad 
\frac{\partial^2 \chi}{\partial t \partial u } =0,\label{mixed0}
\end{equation}
with the additional constraint imposed by the dilatation Ward identity. In particular, in this limit, the dilatation WI for the $(s,t,u)$ invariants takes the form
\begin{align}
\left[s\,\frac{\partial}{\partial s}+t\,\frac{\partial}{\partial t}+u\,\frac{\partial}{\partial u}\right]\chi(s,t,u)=0.\label{dilstu}
\end{align}
Notice that the remaining contribution to the dilatation WI is satisfied separately by the scale invariant condition on $\Phi(p_2,p_3,p_4)$
\begin{equation}
\left[p_2\,\frac{\partial}{\partial p_2}+p_3\,\frac{\partial}{\partial p_3}+p_4\,\frac{\partial}{\partial p_4}\right]\Phi(p_2,p_3,p_4)=(\Delta_t -3 d -2)\Phi(p_2,p_3,p_4),
\end{equation}
which takes to generalized hypergeometric $F_4$ solutions, functions of the ratios $p_2^2/p_4^2$ and $p_3^2/p_4^2$, as given in \eqref{compact}. The choice of the pivot ($p_4$ in this case) is arbitrary.\\
By differentiating \eqref{dilstu} with respect $s$ and using \eqref{mixed0}, one finds another constraint. Similar constraints are obtained by repeating the procedure with respect to $t$ and $u$. The resulting three equations obtained in this manner can be written in the form
\begin{align}
\left[s\frac{\partial^2}{\partial \,s^2}+\,\frac{\partial}{\partial s}\right]\chi(s,t,u)&=0\notag\\
\left[\,t\,\frac{\partial^2}{\partial\,t^2}+\,\frac{\partial}{\partial t}\right]\chi(s,t,u)&=0\notag\\
\left[u\frac{\partial^2}{\partial u^2}+\frac{\partial}{\partial u}\right]\chi(s,t,u)&=0,
\end{align}
giving the solution for $\chi$ of the form
\begin{align}
\chi(s,t,u)=c_1\log(s)+c_2\log(t)+c_3\log(u)+c_4,
\end{align}
where $c_1,\,c_2,\,c_3,\,c_4$ are undetermined constants. Imposing the dilatation WI on this solution we find some relations between the undetermined coefficients, with the solution rewritten in the form
\begin{equation}
\chi(s,t,u)=c_1\log\left(\frac{s}{t}\right)+c_2\log\left(\frac{u}{t}\right)+c_4.
\end{equation}
Finally, we also impose the symmetry constraint on the form factor $A$ of the $TOOO$
\begin{equation}
A(p_2,p_4,p_3,s,u,t)=A(p_2,p_3,p_4,s,t,u)
\end{equation}
which implies that 
\begin{align}
\phi(p_2,p_3,p_4)\chi(s,t,u)=\phi(p_2,p_4,p_3)\chi(s,u,t),
\end{align}
and recalling that the $\phi(p_2,p_3,p_4)$ is symmetric under the permutation of $\{p_2,p_3,p_4\}$, we obtain the condition $\chi(s,t,u)=\chi(s,u,t)$ or
\begin{align}
\big(2c_2+c_1\big)\log\left(\frac{u}{t}\right)=0.
\end{align}
Therefore the $\chi(s,t,u)$ function acquires the final form
\begin{equation}
\chi(s,t,u)=c_1\log\left(\frac{u\,t}{s^2}\right)+c_4.
\end{equation}
As we have seen from the last and the previous cursory analysis of such systems, it is possible to identify an approximate behaviour of such solutions, in one specific asymptotic limit in which 
the invariant $s$ and $t$ get large and of the same size. \\
In this approximate analysis the only exact statement that one can make is that Lauricella functions are indeed special 
solutions of such equations, and correspond to particular solutions of such inhomogenous systems. \\
We have been careful to rewrite all the CWI's for generic scalings $\Delta_i$, in such as way that the left hand sides of thse systems carry a close resemblance to those of 3-point functions, except for an extra term proportional to a double derivative in $s$ and $t$,  
$\sim 1/(s t)\partial^2/(\partial s\partial t)$, which is new for 4-point functions and absent in 3-point functions. \\
As we have stressed in the previous sections in the case of dcc solutions, this term does preserve the hypergeometric structure of the corresponding equations, although such solutions have little in common with those derived for genuine 3-point functions, for being quartic -rather than quadratic - ratios of momenta.  \\
The discovery of such solutions may not be accidental in the context of CFT's, since in ordinary perturbation theory similar dependences have been uncovered in the analysis of ladder diagrams \cite{Usyukina:1992jd}. However, one can easily check, following the discussion in \cite{Maglio:2019grh}, that box-like master integrals with propagators raised to generic powers, cannot be special cases of such dcc solutions, except for the ordinary box diagram. On general grounds, one expects that the simplified CWI's, which
are found in the scalar case for the dcc solutions, are related to an underlying Yangian symmetry \cite{Loebbert:2016cdm}, which is manifesting here in a bosonic, non supersymmetric, context.  
In the $TOOO$ such a symmetry, differently from the scalar case, is violated by the presence of a single stress-energy tensor. It could be restored in tensor correlators characterised by a single primary operator, such as the $JJJJ$ or the $TTTT$. We plan to come back to a discussion of this point in the near future. 

\section{Comments and Conclusions}
The investigation of the CWI's of four point functions of a generic CFT in momentum 
space in $d >2$ is a new challenging domain of research,  with the possibility of establishing a direct connection with the analysis of scattering amplitudes in Lagrangian field theories. As in the case of lower point functions, 
one could envision several areas where such studies could find direct physical applications, from cosmology to condensed matter theory \cite{Chernodub:2017jcp,Chernodub:2019tsx}, due to the interplay, in the latter case, of quantum anomalies in transport phenomena.     
These studies need to be accompanied by investigations of the operator product expansion in the same variables, in order to develop a bootstrap program, as in coordinate space. \\
Obviously, while in coordinate space the operatorial expansion is well-behaved at separate spacetime points, in momentum space we gather information on such operators from all the spacetime regions, including those in which the external coordinates of a correlator coalesce. This makes the analysis in momentum space more demanding, and we have to worry about anomalies and address the issue of how to regulate a given theory. \\
It is then natural to advance our knowledge in this area starting from the analysis of simpler correlation functions, the scalar and the tensor/scalar cases being the first on the list.\\
For this reason we have derived the CWI's for the $TOOO$ and discussed their relation to those obtained in the case of 4 scalars. In both cases we have discussed their expressions in various limits, showing the hypergeometric character of the asymptotic solutions, if certain constraints on the scaling dimensions are respected. \\
While, obviously, we 
do not expect that a given correlator can be uniquely identified by these equations, neverthless they constrain quite significantly the structure of the possible solutions. \\
As mentioned, in our analysis we have concentrated on the structure of the equations in several kinematical limits, in order to gather some information about the behaviour of the corresponding solutions. 
In such limits, the differential operators take a simplified but a still nontrivial form. \\
The comparison between the $TOOO$ and scalar cases, allows to uncover some common features of the systems of equations that they need to satisfy.  In this context, of particular significance are those solutions which are dual conformal and conformal at the same time (or dcc solutions), which take a unique expression. Several different ans\"atze  take to the same hypergeometric form of such solutions, which are related by analytic continuations, and, as we have shown, turn useful for their study in specific kinematical limits. For such a 
reason they play a strategic role, since they can be used to investigate the behaviour of scalar 4-point functions in a rather direct way and allow to underscore some similarities between the CWI's both in the tensor and in the scalar contexts. \\
Specific features of such dcc solutions, extracted in several asymptotic limits, are expected to provide some indication on the behaviour of the more general (and unknown) solutions of the equations satisfied by scalar operators - the $OOOO$ for instance - for generic scaling dimensions of the primaries $O$. Both correlators are characterised by a single form factors, allowing particular solutions of Lauricella type. This suggests the presence of a more general underlying hypergeometric structure in such systems of equations. It could be of interest to investigate from a purely mathematical point of view the structure such equations in order to classify the structure of such solutions.\\
Our investigations can be extended in several directions, for instance to the study of the renormalization of the corresponding form factors, which requires a separate investigation, as in the case of 3-point functions \cite{Bzowski:2015pba}. There are also other and quite direct implications of our results and equations for the analysis of the decomposition of such correlators in terms of CP-symmetric (Polyakov) blocks. Indeed the CWI's that we have derived can be applied to constrain the block decomposition \cite{Isono:2018rrb}. We hope to address these issues in a future work.
\vspace{1cm}

\centerline{\bf Acknowledgements} 
C.C. thanks the Institute for Theoretical Physics at ETH Zurich, and in particular Babis Anastasiou and Vittorio Del Duca for hospitality. D.T. would like to thank Fotis Koutroulis and Konstantinos Rigatos for discussions. This work is partially supported by INFN of Italy, Iniziativa Specifica QFT-HEP.

\appendix
\section{Appendix}
We summarize some definitions and relations concerning the special functions and integrals 
introduced above. 3K integrals can be related to linear combinations of 4 hypergeometric functions 
\begin{align}
& \int_0^\infty d x \: x^{\alpha - 1} K_\lambda(a x) K_\mu(b x) K_\nu(c x) =\frac{2^{\alpha - 4}}{c^\alpha} \bigg[ B(\lambda, \mu) + B(\lambda, -\mu) + B(-\lambda, \mu) + B(-\lambda, -\mu) \bigg], \label{3K}
\end{align}
where
\begin{align}
B(\lambda, \mu) & = \left( \frac{a}{c} \right)^\lambda \left( \frac{b}{c} \right)^\mu \Gamma \left( \frac{\alpha + \lambda + \mu - \nu}{2} \right) \Gamma \left( \frac{\alpha + \lambda + \mu + \nu}{2} \right) \Gamma(-\lambda) \Gamma(-\mu) \times \notag\\
& \qquad \times F_4 \left( \frac{\alpha + \lambda + \mu - \nu}{2}, \frac{\alpha + \lambda + \mu + \nu}{2}; \lambda + 1, \mu + 1; \frac{a^2}{c^2}, \frac{b^2}{c^2} \right), \label{3Kplus}
\end{align}
valid for
\begin{equation}
\Re\, \alpha > | \Re\, \lambda | + | \Re \,\mu | + | \Re\,\nu |, \qquad \Re\,(a + b + c) > 0 \nn
\end{equation}
and the Bessel functions $K_\nu$ satisfy the equations 
\begin{align}
\frac{\partial}{\partial p}\big[p^\b\,K_\b(p\,x)\big]&=-x\,p^\b\,K_{\b-1}(p x)\nn
K_{\b+1}(x)&=K_{\b-1}(x)+\frac{2\b}{x}K_{\b}(x). \label{der}
\end{align}

\section{Primary Conformal Ward Identities in \texorpdfstring{$\bar{p}_1$}{}}\label{AppendixB}
Here we present the explicit expressions of the Primary Conformal Ward Identites in the case of $\bar{p}_1$ dependency. The $C_{ij}$ are given by
\begin{align}
C_{11}&=\Bigg[K_2+\frac{p_3^2-p_4^2}{s\,t}\frac{\partial}{\partial s \partial t}-\frac{p_3^2-p_4^2}{s\,u}\frac{\partial}{\partial s \partial u}+\frac{1}{t}\frac{\partial}{\partial t}\left(p_2\frac{\partial}{\partial p_2}+p_3\frac{\partial}{\partial p_3}-p_4\frac{\partial}{\partial p_4}\right)+(d-\Delta)\left(\frac{1}{t}\frac{\partial}{\partial t}+\frac{1}{u}\frac{\partial}{\partial u}\right)\notag\\
&+\frac{1}{u}\frac{\partial}{\partial u}\left(p_2\frac{\partial}{\partial p_2}-p_3\frac{\partial}{\partial p_3}+p_4\frac{\partial}{\partial p_4}\right)+\frac{2p_2^2+p_3^2+p_4^2-s^2-t^2-u^2}{t\,u}\frac{\partial}{\partial t\partial u}\Bigg]\,A(p_2,p_3,p_4,s,t,u)=0\\
C_{12}&=\Bigg[K_2+\frac{p_3^2-p_4^2}{s\,t}\frac{\partial}{\partial s \partial t}-\frac{p_3^2-p_4^2}{s\,u}\frac{\partial}{\partial s \partial u}+\frac{1}{t}\frac{\partial}{\partial t}\left(p_2\frac{\partial}{\partial p_2}+p_3\frac{\partial}{\partial p_3}-p_4\frac{\partial}{\partial p_4}\right)+(d-\Delta)\left(\frac{1}{t}\frac{\partial}{\partial t}+\frac{1}{u}\frac{\partial}{\partial u}\right)\notag\\
&+\frac{1}{u}\frac{\partial}{\partial u}\left(p_2\frac{\partial}{\partial p_2}-p_3\frac{\partial}{\partial p_3}+p_4\frac{\partial}{\partial p_4}\right)+\frac{2p_2^2+p_3^2+p_4^2-s^2-t^2-u^2}{t\,u}\frac{\partial}{\partial t\partial u}\Bigg]\,A(p_3,p_2,p_4,u,t,s)\notag\\
&\hspace{-0.5cm}+\frac{2}{t}\frac{\partial}{\partial t}\bigg(A(p_2,p_3,p_4,s,t,u)+A(p_3,p_2,p_4,u,t,s)\bigg)-\frac{2}{u}\frac{\partial}{\partial u}\bigg(A(p_2,p_3,p_4,s,t,u)+A(p_3,p_2,p_4,u,t,s)\bigg)
\end{align}
\begin{align}
C_{13}&=\Bigg[K_2+\frac{p_3^2-p_4^2}{s\,t}\frac{\partial}{\partial s \partial t}-\frac{p_3^2-p_4^2}{s\,u}\frac{\partial}{\partial s \partial u}+\frac{1}{t}\frac{\partial}{\partial t}\left(p_2\frac{\partial}{\partial p_2}+p_3\frac{\partial}{\partial p_3}-p_4\frac{\partial}{\partial p_4}\right)+(d-\Delta)\left(\frac{1}{t}\frac{\partial}{\partial t}+\frac{1}{u}\frac{\partial}{\partial u}\right)\notag\\
&+\frac{1}{u}\frac{\partial}{\partial u}\left(p_2\frac{\partial}{\partial p_2}-p_3\frac{\partial}{\partial p_3}+p_4\frac{\partial}{\partial p_4}\right)+\frac{2p_2^2+p_3^2+p_4^2-s^2-t^2-u^2}{t\,u}\frac{\partial}{\partial t\partial u}\Bigg]\,A(p_4,p_3,p_2,t,s,u)\notag\\
&\hspace{-0.5cm}-\frac{2}{t}\frac{\partial}{\partial t}\bigg(A(p_2,p_3,p_4,s,t,u)+A(p_4,p_3,p_2,t,s,u)\bigg)+\frac{2}{u}\frac{\partial}{\partial u}\bigg(A(p_2,p_3,p_4,s,t,u)+A(p_4,p_3,p_2,t,s,u)\bigg)
\end{align}
and 
\begin{align}
C_{21}&=\Bigg[K_3+\frac{p_2^2-p_4^2}{t\, u}\frac{\partial}{\partial t \partial u}-\frac{p_2^2-p_4^2}{s\,u}\frac{\partial}{\partial s \partial u}+\frac{1}{t}\frac{\partial}{\partial t}\left(p_2\frac{\partial}{\partial p_2}+p_3\frac{\partial}{\partial p_3}-p_4\frac{\partial}{\partial p_4}\right)+(d-\Delta)\left(\frac{1}{t}\frac{\partial}{\partial t}+\frac{1}{s}\frac{\partial}{\partial s}\right)\notag\\
&+\frac{1}{s}\frac{\partial}{\partial s}\left(p_3\frac{\partial}{\partial p_3}+p_4\frac{\partial}{\partial p_4}-p_2\frac{\partial}{\partial p_2}\right)+\frac{2p_3^2+p_2^2+p_4^2-s^2-t^2-u^2}{s\,t}\frac{\partial}{\partial s\partial t}\Bigg]\,A(p_2,p_3,p_4,s,t,u)\notag\\
&\hspace{-0.5cm}+\frac{2}{t}\frac{\partial}{\partial t}\bigg(A(p_2,p_3,p_4,s,t,u)+A(p_3,p_2,p_4,u,t,s)\bigg)-\frac{2}{s}\frac{\partial}{\partial s}\bigg(A(p_2,p_3,p_4,s,t,u)+A(p_3,p_2,p_4,u,t,s)\bigg)\\
C_{22}&=\Bigg[K_3+\frac{p_2^2-p_4^2}{t\, u}\frac{\partial}{\partial t \partial u}-\frac{p_2^2-p_4^2}{s\,u}\frac{\partial}{\partial s \partial u}+\frac{1}{t}\frac{\partial}{\partial t}\left(p_2\frac{\partial}{\partial p_2}+p_3\frac{\partial}{\partial p_3}-p_4\frac{\partial}{\partial p_4}\right)+(d-\Delta)\left(\frac{1}{t}\frac{\partial}{\partial t}+\frac{1}{s}\frac{\partial}{\partial s}\right)\notag\\
&+\frac{1}{s}\frac{\partial}{\partial s}\left(p_3\frac{\partial}{\partial p_3}+p_4\frac{\partial}{\partial p_4}-p_2\frac{\partial}{\partial p_2}\right)+\frac{2p_3^2+p_2^2+p_4^2-s^2-t^2-u^2}{s\,t}\frac{\partial}{\partial s\partial t}\Bigg]\,A(p_3,p_2,p_4,u,t,s)\\
C_{23}&=\Bigg[K_3+\frac{p_2^2-p_4^2}{t\, u}\frac{\partial}{\partial t \partial u}-\frac{p_2^2-p_4^2}{s\,u}\frac{\partial}{\partial s \partial u}+\frac{1}{t}\frac{\partial}{\partial t}\left(p_2\frac{\partial}{\partial p_2}+p_3\frac{\partial}{\partial p_3}-p_4\frac{\partial}{\partial p_4}\right)+(d-\Delta)\left(\frac{1}{t}\frac{\partial}{\partial t}+\frac{1}{s}\frac{\partial}{\partial s}\right)\notag\\
&+\frac{1}{s}\frac{\partial}{\partial s}\left(p_3\frac{\partial}{\partial p_3}+p_4\frac{\partial}{\partial p_4}-p_2\frac{\partial}{\partial p_2}\right)+\frac{2p_3^2+p_2^2+p_4^2-s^2-t^2-u^2}{s\,t}\frac{\partial}{\partial s\partial t}\Bigg]\,A(p_4,p_3,p_2,t,s,u)\notag\\
&\hspace{-0.5cm}+\frac{2}{s}\frac{\partial}{\partial s}\bigg(A(p_4,p_3,p_2,t,s,u)+A(p_3,p_2,p_4,u,t,s)\bigg)-\frac{2}{t}\frac{\partial}{\partial t}\bigg(A(p_4,p_3,p_2,t,s,u)+A(p_3,p_2,p_4,u,t,s)\bigg)
\end{align}
and finally 
\begin{align}
C_{31}&=\Bigg[K_4+\frac{p_2^2-p_3^2}{t\, u}\frac{\partial}{\partial t \partial u}-\frac{p_2^2-p_3^2}{s\,t}\frac{\partial}{\partial s \partial t}+\frac{1}{u}\frac{\partial}{\partial u}\left(p_2\frac{\partial}{\partial p_2}-p_3\frac{\partial}{\partial p_3}+p_4\frac{\partial}{\partial p_4}\right)+(d-\Delta)\left(\frac{1}{u}\frac{\partial}{\partial u}+\frac{1}{s}\frac{\partial}{\partial s}\right)\notag\\
&+\frac{1}{s}\frac{\partial}{\partial s}\left(p_3\frac{\partial}{\partial p_3}+p_4\frac{\partial}{\partial p_4}-p_2\frac{\partial}{\partial p_2}\right)+\frac{2p_4^2+p_2^2+p_3^2-s^2-t^2-u^2}{s\,u}\frac{\partial}{\partial s\partial u}\Bigg]\,A(p_2,p_3,p_4,s,t,u)\notag\\
&\hspace{-0.5cm}-\frac{2}{s}\frac{\partial}{\partial s}\bigg(A(p_2,p_3,p_4,s,t,u)+A(p_4,p_3,p_2,t,s,u)\bigg)+\frac{2}{u}\frac{\partial}{\partial u}\bigg(A(p_2,p_3,p_4,s,t,u)+A(p_4,p_3,p_2,t,s,u)\bigg)
\end{align}
\begin{align}
C_{32}&=\Bigg[K_4+\frac{p_2^2-p_3^2}{t\, u}\frac{\partial}{\partial t \partial u}-\frac{p_2^2-p_3^2}{s\,t}\frac{\partial}{\partial s \partial t}+\frac{1}{u}\frac{\partial}{\partial u}\left(p_2\frac{\partial}{\partial p_2}-p_3\frac{\partial}{\partial p_3}+p_4\frac{\partial}{\partial p_4}\right)+(d-\Delta)\left(\frac{1}{u}\frac{\partial}{\partial u}+\frac{1}{s}\frac{\partial}{\partial s}\right)\notag\\
&+\frac{1}{s}\frac{\partial}{\partial s}\left(p_3\frac{\partial}{\partial p_3}+p_4\frac{\partial}{\partial p_4}-p_2\frac{\partial}{\partial p_2}\right)+\frac{2p_4^2+p_2^2+p_3^2-s^2-t^2-u^2}{s\,u}\frac{\partial}{\partial s\partial u}\Bigg]\,A(p_3,p_2,p_4,u,t,s)\notag\\
&\hspace{-0.5cm}+\frac{2}{s}\frac{\partial}{\partial s}\bigg(A(p_3,p_2,p_4,u,t,s)+A(p_4,p_3,p_2,t,s,u)\bigg)-\frac{2}{u}\frac{\partial}{\partial u}\bigg(A(p_3,p_2,p_4,u,t,s)+A(p_4,p_3,p_2,t,s,u)\bigg)\notag\\
\end{align}
\begin{align}
C_{33}&=\Bigg[K_4+\frac{p_2^2-p_3^2}{t\, u}\frac{\partial}{\partial t \partial u}-\frac{p_2^2-p_3^2}{s\,t}\frac{\partial}{\partial s \partial t}+\frac{1}{u}\frac{\partial}{\partial u}\left(p_2\frac{\partial}{\partial p_2}-p_3\frac{\partial}{\partial p_3}+p_4\frac{\partial}{\partial p_4}\right)+(d-\Delta)\left(\frac{1}{u}\frac{\partial}{\partial u}+\frac{1}{s}\frac{\partial}{\partial s}\right)\notag\\
&+\frac{1}{s}\frac{\partial}{\partial s}\left(p_3\frac{\partial}{\partial p_3}+p_4\frac{\partial}{\partial p_4}-p_2\frac{\partial}{\partial p_2}\right)+\frac{2p_4^2+p_2^2+p_3^2-s^2-t^2-u^2}{s\,u}\frac{\partial}{\partial s\partial u}\Bigg]\,A(p_4,p_3,p_2,t,s,u).
\end{align}
\section{Primary Conformal Ward Identities in \texorpdfstring{$\bar{p}_4$}{}}\label{AppendixC}
We present the remaining Primary CWI's of \secref{PCWI22}. We obtain
\begin{align}
\tilde{C}_{12}=&\Bigg[\frac{\partial^2}{\partial p_4^2}+\frac{d-2\Delta+1}{p_4}\frac{\partial}{\partial p_4}-\frac{\partial^2}{\partial p_1^2}-\frac{1-d}{p_1}\frac{\partial}{\partial p_1}+\frac{1}{s}\frac{\partial}{\partial s}\left(p_4\frac{\partial}{\partial p_4}+p_3\frac{\partial}{\partial p_3}-p_1\frac{\partial}{\partial p_1}-p_2\frac{\partial}{\partial p_2}\right)\notag\\&
+\frac{d-\Delta+2}{s}\frac{\partial}{\partial s}
+\frac{p_3^2-p_2^2}{st}\frac{\partial^2}{\partial s \partial t}\Bigg]F(p_1,p_4,p_3,p_2,t,s),\notag \\
\end{align}
\begin{align}
\tilde{C}_{13}=&\Bigg[\frac{\partial^2}{\partial p_4^2}+\frac{d-2\Delta+1}{p_4}\frac{\partial}{\partial p_4}-\frac{\partial^2}{\partial p_1^2}-\frac{1-d}{p_1}\frac{\partial}{\partial p_1}+\frac{1}{s}\frac{\partial}{\partial s}\left(p_4\frac{\partial}{\partial p_4}+p_3\frac{\partial}{\partial p_3}-p_1\frac{\partial}{\partial p_1}-p_2\frac{\partial}{\partial p_2}\right)\notag\\&
+\frac{d-\Delta-2}{s}\frac{\partial}{\partial s}
+\frac{p_3^2-p_2^2}{st}\frac{\partial^2}{\partial s \partial t}\Bigg]F(p_1,p_2,p_4,p_3,s,\tilde{u}),\\
\tilde{C}_{22}=&\Bigg[\frac{\partial^2 }{\partial p_4^2}+\frac{d-2\Delta+1}{p_4}\frac{\partial }{\partial p_4}-\frac{\partial^2 }{\partial p_2^2}-\frac{d-2\Delta+1}{p_2}\frac{\partial }{\partial p_2}+\frac{1}{s}\frac{\partial}{\partial s}\left(p_3\frac{\partial }{\partial p_3}+p_4\frac{\partial }{\partial p_4}-p_1\frac{\partial }{\partial p_1}-p_2\frac{\partial }{\partial p_2}\right)
\notag
\\&+\frac{\Delta-d}{t}\frac{\partial}{\partial t}+\frac{d-\Delta+2}{s}\frac{\partial}{\partial s}+\frac{1}{t}\frac{\partial}{\partial t}\left( p_1 \frac{\partial }{\partial p_1}+p_4\frac{\partial }{\partial p_4}-p_2\frac{\partial }{\partial p_2}-p_3\frac{\partial }{\partial p_3}\right)
\notag\\&+\frac{p_4^2-p_2^2}{st}\frac{\partial^2}{\partial s \partial t}\Bigg]F(p_1,p_4,p_3,p_2,t,s)+\frac{2}{t}\frac{\partial F(p_1,p_2,p_4,p_3,s,\tilde{u})}{\partial t}-\frac{2}{t}\frac{\partial F(p_1,p_2,p_3,p_4,s,t)}{\partial t},\notag\\
\end{align}
\begin{align}
\tilde{C}_{23}=&\Bigg[\frac{\partial^2 }{\partial p_4^2}+\frac{d-2\Delta+1}{p_4}\frac{\partial }{\partial p_4}-\frac{\partial^2 }{\partial p_2^2}-\frac{d-2\Delta+1}{p_2}\frac{\partial }{\partial p_2}+\frac{1}{s}\frac{\partial}{\partial s}\left(p_3\frac{\partial }{\partial p_3}+p_4\frac{\partial }{\partial p_4}-p_1\frac{\partial }{\partial p_1}-p_2\frac{\partial }{\partial p_2}\right)\notag
\\&+
\frac{\Delta-d+2}{t}\frac{\partial}{\partial t}+\frac{d-\Delta-2}{s}\frac{\partial}{\partial s}+\frac{1}{t}\frac{\partial}{\partial t}\left( p_1 \frac{\partial }{\partial p_1}+p_4\frac{\partial }{\partial p_4}-p_2\frac{\partial }{\partial p_2}-p_3\frac{\partial }{\partial p_3}\right)\notag\\&
+\frac{p_4^2-p_2^2}{st}\frac{\partial^2}{\partial s \partial t}\Bigg]F(p_1,p_2,p_4,p_3,s,\tilde{u}),\notag\\
\end{align}
\begin{align}
\tilde{C}_{32}=&\Bigg[\frac{\partial^2 }{\partial p_4^2}+\frac{d-2\Delta+1}{p_4}\frac{\partial }{\partial p_4}-\frac{\partial^2 }{\partial p_3^2}-\frac{d-2\Delta+1}{p_3}\frac{\partial }{\partial p_3}+\frac{p_1^2-p_2^2}{st}\frac{\partial^2}{\partial s \partial t}+\frac{\Delta-d+2}{t}\frac{\partial}{\partial t}\notag\\&+\frac{1}{t}\frac{\partial}{\partial t}\left(p_1\frac{\partial }{\partial p_1}+p_4\frac{\partial }{\partial p_4}-p_2\frac{\partial }{\partial p_2}-p_3\frac{\partial }{\partial p_3}\right)
\Bigg]F(p_1,p_4,p_3,p_2,t,s),\notag\\
\end{align}
\begin{align}
\tilde{C}_{33}=&\Bigg[\frac{\partial^2 }{\partial p_4^2}+\frac{d-2\Delta+1}{p_4}\frac{\partial }{\partial p_4}-\frac{\partial^2 }{\partial p_3^2}-\frac{d-2\Delta+1}{p_3}\frac{\partial }{\partial p_3}+\frac{p_1^2-p_2^2}{st}\frac{\partial^2}{\partial s \partial t}+\frac{\Delta-d}{t}\frac{\partial}{\partial t}\notag
\\&+\frac{1}{t}\frac{\partial}{\partial t}\left(p_1\frac{\partial }{\partial p_1}+p_4\frac{\partial }{\partial p_4}-p_2\frac{\partial }{\partial p_2}-p_3\frac{\partial }{\partial p_3}\right)-\frac{2}{s}\frac{\partial}{\partial s}
\Bigg]F(p_1,p_2,p_4,p_3,t,\tilde{u})\notag\\&
+\frac{2}{s}\frac{\partial F(p_1,p_2,p_3,p_4,s,t)}{\partial s}-\frac{2}{s}\frac{\partial F(p_1,p_4,p_3,p_2,t,s)}{\partial s}\notag\\
&+\frac{2}{t}\frac{\partial F(p_1,p_4,p_3,p_2,t,s)}{\partial t}-\frac{2}{t}\frac{\partial F(p_1,p_2,p_3,p_4,s,t)}{\partial t}.
\end{align}

\section{Appendix}
\label{lauri}
\subsection{The hypergeometric system from \eqref{ipergio}}

Rewriting \eqref{ipergio}, with quadratic ratios $x=q_1^2/q_3^2, y=q_2^2/q_3^2$  
for the correlator $\Phi(q_1,q_2,q_3)=\langle OOO\rangle$  with scaling dimensions $\Delta_i=\Delta,\, i=1,2,3$
\begin{equation}
K_{13}\Phi=0\qquad K_{23}\Phi=0, \nonumber
\end{equation}
one obtains the system of equations
\begin{eqnarray}
\label{diff}
\begin{cases}
 \bigg[ x(1-x) \frac{\partial^2}{\partial x^2} - y^2 \frac{\partial^2}{\partial y^2} - 2 \, x \, y \frac{\partial^2}{\partial x \partial y} +  \left[ \gamma - (\alpha + \beta + 1) x \right] \frac{\partial}{\partial x}  \\
\hspace{8cm} - (\alpha + \beta + 1) y \frac{\partial}{\partial y}  - \alpha \, \beta \bigg] \Phi(x,y) = 0 \,,  \\
\bigg[ y(1-y) \frac{\partial^2}{\partial y^2} - x^2 \frac{\partial^2}{\partial x^2} - 2 \, x \, y \frac{\partial^2}{\partial x \partial y} +  \left[ \gamma' - (\alpha + \beta + 1) y \right] \frac{\partial}{\partial y}  \\
\hspace{8cm} - (\alpha + \beta + 1) x \frac{\partial}{\partial x}  - \alpha \, \beta \bigg] \Phi(x,y) = 0 \,, 
\end{cases} 
\end{eqnarray}
with parameters $\alpha(a,b),\beta(a,b),\gamma(a,b),\gamma'(a,b)$ given in \eqref{cons2} and \eqref{FuchsianPoint}, which are solved by ans\"atz of the form $x^a y^b G(x,y)$. $G$ is an 
Appell function of type $F_4(\alpha,\beta,\gamma,\gamma',x,y)$, given in \eqref{f4}. Both in the case of quadratic or quartic \eqref{xy} ratios, in the variables $x$ and $y$,  the structure of \eqref{diff} is preserved, with appropriate values of the parameters $\alpha(a,b),\beta(a,b),\gamma(a,b),\gamma'(a,b)$ and indices $a,b$.

\subsection{The Lauricella system and 4K}
In the $s,t \to \infty$ limit the equations for the $K_{ij}$ operators, for arbitrary scalings $\Delta_i$, can be organized in the form

\begin{equation}
\textup{K}_{14}\phi=0,\qquad \textup{K}_{24}\phi=0,\qquad \textup{K}_{34}\phi=0\label{CWILaur}
\end{equation}
where 
\begin{align}
\textup{K}_i&=\frac{\partial^2}{\partial p_i^2}+\frac{(d-2\D_i+1)}{p_i}\frac{\partial}{\partial p_i},\qquad i=1,\dots,4\ ,\\
\textup{K}_{ij}&=\textup{K}_i-\textup{K}_j\ .
\end{align}
One can choose an arbitrary momentum as pivot in the ansatz for the solution of such system, for instance $(x,y,z,p_4^2)$, where 
\begin{equation}
x=\sdfrac{p_1^2}{p_4^2},\quad y=\sdfrac{p_2^2}{p_4^2},\quad z=\sdfrac{p_3^2}{p_4^2}
\end{equation}
are dimensionless quadratic ratios. The ans\"atz for the solution can be taken of the form
\begin{equation}
\phi(p_1,p_2,p_3,p_4)=(p_4^2)^{n_s}\,x^a\,y^b\,z^c\,F(x,y,z),
\end{equation}
satisfying the dilatation Ward identity with the condition
\begin{equation}
n_s=\frac{\D_t}{2}-\frac{3d}{2} \qquad \Delta_t=\sum_{i=1}^4 \Delta_i.
\end{equation}
With this ans\"atz the conformal Ward identities takes the form
\begin{align}
\textup{K}_{14}\phi=&4p_4^{\D_t-3d-2}\,x^a\,y^b\,z^c\,\bigg[(1-x)x\sdfrac{\partial^2}{\partial x^2}-2x\,y\sdfrac{\partial^2}{\partial x\partial y}-y^2\sdfrac{\partial^2}{\partial y^2}-2x\,z\sdfrac{\partial^2}{\partial x\partial z}-z^2\sdfrac{\partial^2}{\partial z^2}-2y\,z\sdfrac{\partial^2}{\partial y\partial z}\notag\\
&\hspace{2cm}+(Ax+\gamma)\sdfrac{\partial}{\partial x}+Ay\sdfrac{\partial}{\partial y}+Az\sdfrac{\partial }{\partial z}+\left(E+\sdfrac{G}{x}\right)\bigg]F(x,y,z)=0,
\end{align}
with
\begin{subequations}
	\begin{align}
	A&=\D_1+\D_2+\D_3-\sdfrac{5}{2}d-2(a+b+c)-1\\
	E&=-\sdfrac{1}{4}\big(3d-\D_t+2(a+b+c)\big)\big(2d+2\D_4-\D_t+2(a+b+c)\big)\\
	G&=\sdfrac{a}{2}\,\left(d-2\D_1+2a\right)\\
	\g&=\sdfrac{d}{2}-\D_1+2a+1.
	\end{align}
\end{subequations}
Similar constraints are obtained from the equations $\textup{K}_{24}\phi=0$ and $\textup{K}_{34}\phi=0$. \\
The reduction to the hypergeometric form requires that all the $1/x, 1/y$ and $1/z$ terms of the equations vanish. This implies that the Fuchsian points $a,b,c$ have  values 
\begin{subequations}
\label{ind}
	\begin{align}
	a&=0,\,\D_1-\sdfrac{d}{2} \nn
	b&=0,\,\D_2-\sdfrac{d}{2} \nn
	c&=0,\,\D_3-\sdfrac{d}{2}
	\end{align}
\end{subequations}
and 
\begin{align}
\a(a,b,c)&=d+\D_4-\sdfrac{\D_t}{2}+a+b+c\notag\\
\b(a,b,c)&=\sdfrac{3d}{2}-\sdfrac{\D_t}{2}+a+b+c
\end{align}

\begin{equation}
\g(a)=\frac{d}{2}-\Delta_1+2a+1\,,\qquad\g'(b)=\frac{d}{2}-\Delta_2+2b+1\,,\qquad\g''(c)=\frac{d}{2}-\Delta_3+2c+1.
\end{equation}
With this redefinition of the coefficients, the equations are then expressed in the form
\begin{equation}
\resizebox{1\hsize}{!}{$
\left\{
\begin{matrix}
&x_j(1-x_j)\sdfrac{\partial^2F}{\partial x_j^2}+\hspace{-1cm}\sum\limits_{\substack{\hspace{1.3cm}s\ne j\ \text{for}\ r=j}}\hspace{-1.1cm}x_r\hspace{0.2cm}\sum x_s\hspace{0.5ex}\sdfrac{\partial^2F}{\partial x_r\partial x_s}+\left[\g_j-(\a+\b+1)x_j\right]\sdfrac{\partial F}{\partial x_j}-(\a+\b+1)\sum\limits_{k\ne j}\,x_k\sdfrac{\partial F}{\partial x_k}-\a\,\b\,F=0\\[3ex]
& (j=1,2,3)
\end{matrix}\right.\label{systemLauricella}$}
\end{equation}
where we have set $\g_1=\g$, $\ \g_2=\g'$ and $\g_3=\g''$ and $x_1=x$, $x_2=y$ and $x_3=z$. 
The system of equations admits as solutions hypergeometric functions of three variables, the Lauricella functions, of the form
\begin{equation}
F_C(\a,\b,\g,\g',\g'',x,y,z)=\sum\limits_{m_1,m_2,m_3}^\infty\,\frac{(\a)_{m_1+m_2+m_3}(\b)_{m_1+m_2+m_3}}{(\g)_{m_1}(\g')_{m_2}(\g'')_{m_3}m_1!\,m_2!\,m_3!}x^{m_1}y^{m_2}z^{m_3}.
\end{equation}
where the Pochhammer symbol $(\l)_{k}$ with an arbitrary $\l$ and $k$ a positive integer has been defined in \eqref{Pochh}. The convergence region of this series is defined by the condition
\begin{equation}
\left|\sqrt{x}\right|+\left|\sqrt{y}\right|+\left|\sqrt{z}\right|<1.
\end{equation}
The function $F_C$ is the generalization of the Appell $F_4$ for the case of three variables.
The system of equations \eqref{systemLauricella} admits 8 independent particular integrals (solutions) listed below. Finally, the solution for $\phi$  can be written as
\begin{equation}
\phi(p_i^2)=p_4^{\D_t-3d}\sum_{a,b} C_{i \,(a,b)}\ x^a y^b z^c F_C(\a(a,b),\b(a,b),\g(a,b),\g'(as,b),\g''(a,b),x,y,z)
\end{equation}
where $C_i$ are arbitrary constants. The sum runs over all the possible triple $(a,b,c)$ identified in \eqref{ind}. Introducing the 4K integral
\begin{equation}
I_{\a\{\b_1,\b_2,\b_3,\b_4\}}(p_1,p_2,p_3,p_4)=\int_0^\infty\,dx\,x^\a\,\prod_{i=1}^4(p_i)^{\b_i}\,K_{\b_i}(p_i\,x),
\end{equation}
the same solution can be re-expressed in the form 
\begin{align}
\phi(p_1,p_2,p_3,p_4)&=C\, I_{d-1\left\{\D_1-\frac{d}{2},\D_2-\frac{d}{2},\D_3-\frac{d}{2},\D_4-\frac{d}{2}\right\}}(p_1,p_2,p_3,p_4)\notag\\
&=\int_0^\infty\,dx\,x^{d-1}\,\prod_{i=1}^4(p_i)^{\D_i-\frac{d}{2}}\,K_{\D_i-\frac{d}{2}}(p_i\,x),
\label{4Kfin}
\end{align}
where $C$ is a undetermined constant. 
\section{Appendix}
\subsection{DCC solutions}
Dual conformal/conformal correlators, in the case of scalar 4-point functions $(\Phi)$, are defined by the condition that if we redefine in momentum space the momentum dependence in the form 
\begin{equation}
\label{mapp}
p_1=y_{12}\qquad p_2=y_{23} \qquad p_{3}=y_{34} \qquad p_4=y_{41},
\end{equation}
in the dual variable $y_i$ (i=1,2,3,4), with $\Phi(p_i)\to \Phi(y_i)$,  such correlators satisfy the same CWI's as usually defined in the ordinary variables $x_i$. This condition obviously 
constrains the expression of the correlator to take the form (for equal scalings $\Delta$)
\begin{equation}
\Phi(y_i)=\frac{1}{y_{12}^{2\Delta} y_{34}^{2 \Delta}}h(u(y_i),v(y_i)),
\label{based}
\end{equation}
where the two conformal invariant ratios are given by 
\begin{equation} 
u(y_i)=\frac{y_{12}^2 y_{34}^2}{y_{13}^2 y_{2 4}^2}\qquad  v(y_i)=\frac{y_{23}^2 y_{41}^2}{y_{13}^2 y_{2 4}^2},
\end{equation}
giving the quartic ratios defined in \eqref{xy}

 \begin{equation}
 x=\frac{p_1^2 p_3^2}{s^2 t^2}\qquad  y=\frac{p_2^2 p_4^2}{s^2 t^2}.
 \end{equation}
 using the mapping \eqref{mapp}.
At this stage, one can introduce a dual conformal ans\"atz in terms of $x$ and $y$ based on 
\eqref{based}, and impose the condition that \eqref{based} is a solution of the ordinary CWI's in momentum space. These (dcc) conditions take to an hypergeometric solution of the form \eqref{twoo} or, equivalently, to \eqref{fform}. 


\end{document}